\documentclass[journal]{IEEEtai}

\usepackage{cite}
\usepackage{amssymb}
\usepackage{pifont}
\usepackage{algorithmic}
\usepackage{graphicx}
\usepackage{textcomp}
\usepackage{multirow}
\usepackage{xcolor}
\usepackage{bbding}
\usepackage{footnote}
\usepackage{amsmath, amsfonts}
\usepackage{tablefootnote}

\usepackage{algorithm}
\usepackage{hyperref}
\usepackage{float}
\usepackage{xcolor}
\usepackage{subcaption}
\hypersetup{
    colorlinks=true,
    citecolor=black,
    linkcolor=black,
    filecolor=black,      
    urlcolor=black,
}

\usepackage{color,array}

\begin{document}

\title{Retrieval-Augmented Review Generation for Poisoning Recommender Systems}

\author{Shiyi Yang, Xinshu Li, Guanglin Zhou, Chen Wang, Xiwei Xu,~\IEEEmembership{Senior Member,~IEEE}, \\
Liming Zhu,~\IEEEmembership{Senior Member,~IEEE}, 
Lina Yao,~\IEEEmembership{Senior Member,~IEEE}
\thanks{Shiyi Yang is with University of New South Wales and CSIRO's Data61, Sydney, Australia, {\tt\small shiyi.yang1@unsw.edu.au}.}
\thanks{Xinshu Li is with Macquarie University, Sydney, Australia, {\tt\small xinshu.li@mq.edu.au}.}
\thanks{Guanglin Zhou is with University of Queensland, Brisbane, Australia, {\tt\small guanglin.zhou@uq.edu.au}.}
\thanks{Chen Wang, Xiwei Xu, Liming Zhu, Lina Yao are with CSIRO's Data61 and University of New South Wales, Sydney, Australia, {\tt\small chen.wang, xiwei.xu, liming.zhu, lina.yao@data61.csiro.au}.}}

\maketitle

\begin{abstract}
Recent studies have shown that recommender systems (RSs) are highly vulnerable to data poisoning attacks, where malicious actors inject fake user profiles, including a group of well-designed fake ratings, to manipulate recommendations. 
Due to security and privacy constraints in practice, attackers typically possess limited knowledge of the victim system and thus need to craft profiles that have transferability across black-box RSs. To maximize the attack impact, the profiles often remains imperceptible. However, generating such high-quality profiles with the restricted resources is challenging. Some works suggest incorporating fake textual reviews to strengthen the profiles; yet, the poor quality of the reviews largely undermines the attack effectiveness and imperceptibility under the practical setting. To tackle the above challenges, in this paper, we propose to enhance the quality of the review text by harnessing in-context learning (ICL) capabilities of multimodal foundation models. To this end, we introduce a demonstration retrieval algorithm and a text style transfer strategy to augment the navie ICL. Specifically, we propose a novel practical attack framework named RAGAN to generate high-quality fake user profiles, which can gain insights into the robustness of RSs. The profiles are generated by a jailbreaker and collaboratively optimized on an instructional agent and a guardian to improve the attack transferability and imperceptibility. Comprehensive experiments on various real-world datasets demonstrate that RAGAN achieves the state-of-the-art poisoning attack performance.
\end{abstract}

\begin{IEEEImpStatement}
Recommender systems play a vital role across e-commerce, online content, and social media platforms, benefiting both users and businesses through personalized suggestions and improved engagement. These advantages also create incentives for malicious actors to exploit them. Recent studies reveal that modern recommender systems are vulnerable to data poisoning attacks, leading to unfair competition and loss of user trust. However, existing attack methods often have limited practicality, overestimating system robustness under real-world constraints. To address this, we introduce a retrieval-augmented review generation framework that improves both the effectiveness and efficiency of such attacks. Experiments on real-world datasets show that RAGAN uncovers hidden practical vulnerabilities across multiple AI-based recommenders, outperforming state-of-the-art methods by up to 70\% in attack performance while producing more natural review text. These findings offer actionable guidance for developing robust, transparent, and trustworthy recommender systems, supporting safer digital marketplaces and fostering public trust in AI technologies.
\end{IEEEImpStatement}


\section{Introduction} \label{introduction}

Recommender systems (RSs) mitigate information overload by mining users’ historical interactions to help them discover desirable products. At the same time, they benefit businesses by increasing customer retention and boosting sales through product promotion \cite{cohen2021black}. As a result, RSs have been widely deployed in various platforms such as e-commerce (e.g., Amazon, Taobao, Yelp) and social media (e.g., Facebook, YouTube, LinkedIn) \cite{chen2022knowledge}. The great influences of RSs on individuals/organizations and their openness nature \cite{zhang2021data} provide both incentives and conveniences for unscrupulous parties to poison RSs. Specifically, attackers can inject a group of well-crafted fake user profiles into the training data of target RSs, including user-item interaction data, to mislead the RSs for certain malicious purposes (e.g., to promote their own products for profits). This kind of attacks is known as data poisoning attacks \cite{gunes2014shilling}. 
Many academic studies \cite{si2020shilling} and practical applications \cite{lin2022shilling} have shown that real-world RSs are highly vulnerable to such attacks. The attacks undermine users' trust in RSs and lead to unfair competition among businesses.

To proactively identify potential risks before real attacks, gain insight into the robustness of existing RSs, and drive the development of effective defensive measures, increasing efforts have been devoted to studying how RSs can be attacked \cite{gunes2014shilling, li2016data, yang2017fake, fang2018poisoning, christakopoulou2019adversarial, lin2020attacking, si2020shilling, fang2020influence, tang2020revisiting, fan2021attacking, zhang2021attacking, huang2021data, zhang2021data, chen2021data, cohen2021black, wu2021triple, lin2022shilling, chen2022knowledge, zeng2023practical, chiang2023shilling}. However, existing attack methods are \textit{still far from practical}, which may \textit{overestimate} the robustness of RSs in realistic scenarios \cite{zeng2023practical}. As discussed in previous works \cite{yang2023incorporated}, a practical attack should meet three criteria: 1) {\textit{Knowledge-restricted settings:}} In real-world RSs, attackers usually have limited knowledge of the target systems, including their architectures, parameters, and training data, due to system complexity, frequent updates, and privacy protections \cite{lin2020attacking, lin2022shilling, zhang2021data}; 2) {\textit{Transferability:}} Since the actual target RS is typically treated as a black-box \cite{zeng2023practical}, fake user profiles generated by the attack should have transferability (i.e., effective against different black-box RSs and training data); and 3) {\textit{Imperceptibility:}} Malicious profiles are usually designed to closely resemble legitimate ones \cite{zhang2021attacking}, making them more difficult to detect and hence extending both the duration and scope of their impact in practice.

Existing studies satisfy only a subset of the above outlined criteria, preventing them from forming a comprehensive and robust solution. Conventional poisoning methods \cite{mobasher2007attacks, li2016data, huang2021data} primarily focus on optimizing attack effectiveness for specific collaborative filtering (CF) models \cite{zhang2019deep}, but often neglect stealthiness, making them more detectable. To mitigate this, recent studies have leveraged generative adversarial networks (GANs) \cite{goodfellow2020generative} to produce more inconspicuous perturbations. However, excessive emphasis on imperceptibility may compromise the transferability of attacks \cite{christakopoulou2019adversarial}. To address this trade-off, subsequent works \cite{lin2020attacking, lin2022shilling} have integrated transferability optimization into the GAN framework, thereby improving both objectives simultaneously. By producing profiles containing only fake numerical ratings, these attacks have semantic gaps with contemporary RSs enhanced by textual reviews \cite{sachdeva2020useful}. To solve this, a prior work, R-Trojan \cite{yang2023incorporated}, proposes to incorporate textual reviews, which is a publicly available natural feature and contains rich semantic information reflecting user behaviors and item characteristics \cite{zheng2017joint}, to reinforce attack profiles. Nevertheless, the reviews generated by R-Trojan are of low quality, characterized by logical gaps, incoherence, factual inaccuracies, sparse content, and so on, rendering them noticeable and weakening the effectiveness of transferability optimization. Moreover, R-Trojan makes an unrealistic assumption that the full data can be accessed.

\begin{savenotes}
\begin{table*}
  \caption{Comparison of data poisoning attack approaches, where $\mathcal{|U|}$ and $\mathcal{|V|}$ denote the number of users and items, respectively, $l$ is the length of recommendation list and $p$ represents the maximum percentage of the data that the attack requires. Note that several methods (e.g., TrialAttack and RegUP) propose multiple attacks that require different levels of knowledge of the target RS, and we choose the attacks require the least knowledge for comparison due to they are closest to real-world settings. Note that \checkmark~= available; $\times$~= not available.}
  \vspace{-1em}
  \begin{center}
  \small
  \resizebox{2\columnwidth}{!}{
  \label{tab:attack_works}
  \begin{tabular}{cc|ccc|cc|c}
    \hline
     \multicolumn{2}{c|}{\multirow{3}{*}{Attack Method}} &  \multicolumn{3}{c|}{Knowledge of Target Recommendation Systems} & \multicolumn{2}{c|}{Attack Objectives} & \multirow{3}{*}{Multimodal Profiles?}\\
    
    \cline{3-7}
    & &  \multirow{2}{*}{Training Data} & \multirow{2}{*}{RS Parameters} & \multirow{2}{*}{RS Architectures} & Discuss transferability & Discuss imperceptibility\\

    & &  &  &  &  or not? & or not? \\

    \hline
    
     \multicolumn{1}{c|}{\multirow{2}{*}{Conventional}} & Random &   $\mathcal{|U|} \cdot \mathcal{|V|}$  & $\times$ & $\times$ & $\times$ & $\times$ &  $\times$\\
     \multicolumn{1}{c|}{} & Bandwagon &  $\mathcal{|U|} \cdot \mathcal{|V|}$  & $\times$ & $\times$ & $\times$ & $\times$ &  $\times$\\

 \hline

    \multicolumn{1}{c|}{\multirow{5}{*}{Algorithm-Specific}} & MF-based & $\mathcal{|U|} \cdot \mathcal{|V|}$ & \checkmark  & \checkmark  & $\times$ & $\times$ &  $\times$\\
    \multicolumn{1}{c|}{} & AR-based & $l \cdot \mathcal{|V|}$  &  $\times$ & $\times$ &  \checkmark  & $\times$ &  $\times$\\
    \multicolumn{1}{c|}{} & Graph-based & $\mathcal{|U|} \cdot \mathcal{|V|}$ &  $\times$   & $\times$ & \checkmark & \checkmark &  $\times$\\
    \multicolumn{1}{c|}{} & DL-based & $\mathcal{|U|} \cdot \mathcal{|V|}$  & $\times$   & \checkmark  & \checkmark & \checkmark &  $\times$\\

    \multicolumn{1}{c|}{} & Review-based & $\mathcal{|U|} \cdot \mathcal{|V|}$  & $\times$   & $\times$ & \checkmark & \checkmark &  $\times$\\

     \hline
    
     \multicolumn{1}{c|}{\multirow{7}{*}{GAN-based}} & DCGAN & $\mathcal{|U|} \cdot \mathcal{|V|}$  & $\times$   & $\times$& $\times$  & \checkmark &  $\times$\\
     \multicolumn{1}{c|}{} & AUSH & $\mathcal{|U|} \cdot \mathcal{|V|}$  & $\times$   & $\times$  & \checkmark & \checkmark &  $\times$\\
    \multicolumn{1}{c|}{} & TrialAttack  & $p \cdot \mathcal{|U|} \cdot \mathcal{|V|}$ & $\times$   &  \checkmark & $\times$ & \checkmark &  $\times$\\
     \multicolumn{1}{c|}{} & RecUP & $p \cdot \mathcal{|U|} \cdot \mathcal{|V|}$  & $\times$   & $\times$ & $\times$  & \checkmark &  $\times$ \\
     \multicolumn{1}{c|}{} & Leg-UP & $\mathcal{|U|} \cdot \mathcal{|V|}$ & $\times$   & $\times$ & \checkmark  & \checkmark &  $\times$ \\
      \multicolumn{1}{c|}{} & R-Trojan & $\mathcal{|U|} \cdot \mathcal{|V|}$ &   $\times$ & $\times$   & \checkmark  & \checkmark &  \checkmark \\
       \multicolumn{1}{c|}{} & RAGAN & $p \cdot \mathcal{|U|} \cdot \mathcal{|V|}$ &   $\times$ & $\times$   & \checkmark  & \checkmark  &  \checkmark \\
 \hline
  \end{tabular}}
  \end{center}
\vspace{-1.8em}
\end{table*}
\end{savenotes}

To tackle the above challenges, we propose a novel practical attack framework named RAGAN for generating high-quality fake user profiles within the limited accessible resources, including partial data. The attack profiles are constructed with numerical ratings and enhanced with semantically meaningful textual reviews. These reviews provide plausible justifications for the ratings, enhancing the imperceptibility of the actual attack, and convey persuasive content that influences user decisions, thereby improving the real attack’s effectiveness.

RAGAN is a tailored GAN improved by multimodal foundation models (FMs), composed of a jailbreaker, an instructional agent and a guardian. The jailbreaker aims to produce high-quality attack profiles, where the ratings are produced through pattern learning, and the corresponding reviews are generated by harnessing in-context learning (ICL) capabilities of multimodal foundation models \cite{zhou2024adapting, luo2024context}. To improve the quality of the reviews to strengthen the profiles' quality, we introduce a demonstration retrieval algorithm and a text style transfer strategy to augment the naive ICL. The instructional agent is introduced to optimize the profiles to improve the transferability of the attack by providing constructive feedback to the jailbreaker. To enhance the accuracy of the feedback under the limited knowledge, we adopt review-based RSs for building agents, due to review text can finely reflect user behaviors and characteristics of their preferred items, as compared to rating-only-based RSs \cite{lin2022shilling, zhao2024recommender}. Similarly, the guardian is a review-based detector, which is designed to distinguish the generated attack profiles from the benign profiles as much as possible for optimizing the attack imperceptibility.

Our main contributions are summarized as follows:

\begin{itemize}
    \item We propose a new attack framework named RAGAN, which can generate transferable and imperceptible fake user profiles under a practical setting.

    \item We propose a demonstration retrieval algorithm and a text style transfer strategy to harness the in-context learning for high-quality attack profile generation.

    \item We comprehensively demonstrate the effectiveness of semantically rich textual reviews in improving the quality of attack profiles, even within the limited resources.
    
    \item Extensive experimental results on real-world datasets show that RAGAN significantly outperforms state-of-the-art attack methods in terms of transferability and imperceptibility.
\end{itemize}

\section{Related Works} \label{related_work}

RSs are highly vulnerable to data poisoning attacks, aka shilling attacks, due to their openness  (i.e., the training data of RSs is usually publicly accessible) \cite{si2020shilling, wang2024poisoning}. Researchers have successfully performed shilling attacks against real-world RS such as YouTube, Google Search, Amazon and Yelp in experiments \cite{lin2022shilling}. Large companies (e.g., Amazon, Sony, eBay) have also reported that they have suffered from such attacks in practice \cite{lin2022shilling, tang2020revisiting}. In poisoning attacks, a number of well-crafted fake user profiles can be injected to the victim RS to promote/demote an attacker-chosen item. Considering that demoting a target item can be implemented by promoting other items \cite{huang2021data}, current works focus mainly on item promotions for simplicity \cite{wang2024poisoning}. The items in an attack profile have four different types \cite{gunes2014shilling}: selected items, filler items, unrated items and the target item. To be specific, the selected items are determined by the attacker and a rating function that reflect the characteristics of the attack, and are optionally included in the profiles. The filler items make the fake user profile resemble the real user profile, reducing the chance of being detected. The target item is chosen by the attacker for promotion or demotion, and the remaining items are unrated. Based on the modality of the generated profiles, attacks can be categorized into two types: 1) \textit{unimodal attacks} that involve only numerical ratings or textual reviews, and 2) \textit{multimodal attacks} that incorporate both ratings and review text, leading to higher-quality profiles with enhanced realism and manipulative power.

\subsection{Unimodal Attacks on Recommender Systems}

Traditional RSs often use CF methods such as KNN \cite{bell2007improved}, AR \cite{lee2001web} and MF \cite{koren2009matrix} to model user-item interaction matrix and predict users' preference on items, where MF has been widely adopted because of its performance and flexibility \cite{zhang2019deep}. 
Conventional shilling attacks \cite{gunes2014shilling} such as Random and Bandwagon \cite{mobasher2007attacks} 
rely on global statistics and work mainly for the traditional CFs (e.g., user-based KNN) \cite{lin2022shilling}. These methods are simple heuristics based and not transferable among different RSs \cite{lin2020attacking} (e.g., for item-based KNN \cite{mobasher2007toward}). The lack of diversity in data generation makes them easy to be detected \cite{lin2022shilling}.

Some data poisoning attacks are then proposed to optimize for specific types of RSs, such as PGA \cite{li2016data} for the MF-based, Co-visitation Injection \cite{yang2017fake} for the AR-based and a black-box poisoning strategy designed for the graph-based \cite{fang2018poisoning}. These attacks typically outperform conventional shilling attacks on the RSs they are tailored for. As shown in Table~\ref{tab:attack_works}, \cite{yang2017fake} and \cite{fang2018poisoning} explore the transferability of poisoning attacks, with a focus on the effectiveness of the attacks among traditional black-box RSs. Traditional CF methods are gradually replaced by non-linear deep learning (DL) paradigms \cite{zhang2019deep}, due to their restricted abilities to capture complex structure of massive interaction data. The typical approaches are NCF \cite{he2017neural} and LightGCN \cite{he2020lightgcn}. However, there is a gap between existing traditional algorithm-specific attacks \cite{li2016data, yang2017fake, fang2018poisoning} and the growing prevalence of DL-based paradigms. To fill this gap, Huang et al. \cite{huang2021data} propose a data poisoning attack tailored for DL-based RSs. Nevertheless, they demonstrate the transferability of the attack under unrealistic assumptions, as shown in Table \ref{tab:attack_works}.

Textual reviews, as a natural source of data containing rich semantic information, have been increasingly integrated into DL-based RSs to address the data sparsity and cold-start issues, enhance recommendation accuracy and improve model interpretability \cite{sachdeva2020useful, xu2021understanding}. A representative example is DeepCoNN \cite{zheng2017joint}, which jointly models user and item reviews to learn more expressive latent representations. Chiang et al. \cite{chiang2023shilling} fine-tune GPT-2 \cite{brown2020language} through reinforcement learning  to generate fake textual reviews for prediction shifts in review-based RSs. However, their method 1) requires that the feedback from RSs is periodically available, which  is impractical for most RSs \cite{lin2022shilling}; and 2) generates the review-only profiles, which is not applicable to rating-based DL methods. Since we focus on promoting attacks aligned with Top-N recommendation scenarios prevalent in real-world applications, such a shilling attack targeting rating predictions and inducing prediction shift falls outside the scope of this paper. In addition, these algorithm-specific attacks focus solely on attack performance optimization, making their imperceptibility limited.

Recently, some efforts have been made to leverage GAN for effective profile generation, e.g., DCGAN \cite{christakopoulou2019adversarial}, AUSH \cite{lin2020attacking}, TrialAttack \cite{wu2021triple}, RecUP \cite{zhang2021attacking} and Leg-UP \cite{lin2022shilling}, as illustrated in the table. DCGAN, TrialAttack and RecUP do not discuss the effectiveness of attacks on different black-box RSs.
As pointed out in \cite{zhang2021data}, it is impractical to assume that the entire training data of the target RS can be obtained by the attacker in the real world. The possible reasons are 1) service providers may restrict one’s access to full data or even perturb these data to enhance user privacy and 2) the attacker’s ability to collect massive data may be limited by his resources and/or the platform’s data volume restriction. Consequently, these issues may make AUSH and Leg-UP suffer transferability degradation in practice due to their full data assumptions. In addition, these GAN-based methods are unable to transferable to review-based RSs, and are overestimated detection escape capabilities by assuming that all the data are available.

\subsection{Multimodal Attacks on Recommender Systems}

R-Trojan \cite{yang2023incorporated} has been proposed to mitigate the most above problems, which is a tailored transformer-powered GAN for generating attack profiles containing not only numerical ratings but also textual reviews. However, as illustrated in Table~\ref{tab:attack_works}, R-Trojan makes an unrealistic assumption of full access to the learning data, and relies on fine-tuning equipped with very simple prompts (i.e., topics + sentiments) to generate fake textual reviews. These design choices make R-Trojan computationally expensive and impractical when model parameters are inaccessible (for instance, with state-of-the-art closed models such as GPT-5). In addition, the generated text often suffers from issues such as factual inaccuracy, repetitive phrasing, semantic confusion and logical inconsistencies \cite{brown2020language}, leading to the low-quality profiles and thus limited transferability and imperceptibility. To solve the problems, RAGAN introduces a more efficient and effective generation scheme. By harnessing ICL capabilities of multimodal FMs \cite{chua2024ai} that pre-trains on extensive and diverse data and is built upon flexible and advanced transformers, it produces reviews that are more semantically meaningful, coherent and natural, thereby improving the transferability and imperceptibility of attack profiles under a more practical setting (i.e., beyond black-box RSs, the setting involves partial data). Crucially, ICL enables flexible and scalable adaptation without any additional training or parameter updates, in contrast to fine-tuning.

\section{Problem Formulation} \label{problem_formualtion}
In this section, we first introduce necessary preliminaries and notations used for the following works, then present our threat model, and finally formulate our poisoning attack RAGAN as a bi-level optimization problem.

\subsection{Preliminaries and Notations}
We use $\mathcal{M} = \{m_{u,v}: u \in \mathcal{U}, v \in \mathcal{V}\}$ to denote the records of the user-item interaction matrix in the large-scale recommendation space, where $\mathcal{U}$ and $\mathcal{V}$ represent the set of real users and the item universe, respectively. An entry $m_{u,v}$ in $\mathcal{M}$ is $(r_{u,v}, \delta_{u,v})$, in which $u$ is the row No. and $v$ is the column No. $r_{u, v}$ is the numerical rating from user $u$ on item $v$ and $\delta_{u, v}$ is the corresponding review text. $\mathcal{V}_u = \{v \in \mathcal{V}: r_{u,v} \neq 0\}$ indicates the set of items that have been interacted by $u$ (i.e., the user's profile). Similarly, $\mathcal{U}_v = \{u \in \mathcal{U}: r_{u,v} \neq 0\}$ represents the set of users that have rated $v$. Similar to the benign matrix $\mathcal{M}$, we adopt $\mathcal{\widetilde{M}}  = \{(r_{\widetilde{u},v}, \delta_{\widetilde{u},v}): \widetilde{u} \in \widetilde{\mathcal{U}}, v \in \mathcal{V}\}$ to indicate the fake matrix, where $\widetilde{\mathcal{U}}$ is the fake user pool, and $r_{\widetilde{u},v}$ and $\delta_{\widetilde{u},v}$ are, respectively, fake numerical ratings and fake textual reviews that will be generated and optimized by RAGAN. Thus, each row of $\mathcal{\widetilde{M}}$ is a fake user profile. The target item is represented by $t$.
Moreover, $\mathcal{D}=\{(x_v, d_v, n_v): v \in \mathcal{V}\}$ is used to denote the raw metadata of the items, where $x_v$ is the image of the item $v$, $d_v$ is the textual description of the item $v$, $n_v$ is the title of the item $v$.

\subsection{Threat Model} \label{threat_model}

\subsubsection{\textbf{Attacker's Objective}} 
As discussed in Section \ref{introduction}, we consider two practical properties of an RS attack method.
\begin{enumerate}
    \item    
    \textit{Transferability.} Let $h(t)$ to denote the hit ratio of a target item $t$, which is the percentage of normal users whose the recommendation lists include $t$ after the attack, i.e., $h(t) = \sum_{u\in \widehat{\mathcal{U}}_t} \frac{\mathbb{I}(t \in \widehat{ \mathcal{V}}_{u,:k})}{|\widehat{\mathcal{U}}_t|}$, where $\mathbb{I}(\cdot)$ is an indicator function, $\widehat{\mathcal{U}}_t = \{u \in \mathcal{U}: r_{u,t} = 0\}$ is the set of normal users (i.e., target users who have not yet interacted with $t$) and $\widehat{\mathcal{V}}_{u,:k}$ is the top-$k$ recommendation list of the normal user $u$. Our primary objective is to maximize $h(t)$ on the black-box RSs.

    \item \textit{Imperceptibility.} Our secondary attack objective is to make our attack as imperceptible as possible to maximize the attack impact and the number of affected users.

\end{enumerate}

\subsubsection{\textbf{Attacker's Knowledge}} We assume that the attacker is able to access to \textit{a fraction $p$ of the training data $\mathcal{M}$ of the victim RS}, where $p=1$ denotes full knowledge of data and $0<p<1$ indicates partial knowledge of data.
This is realistic considering the openness nature  \cite{si2020shilling} of a RS.
Although our attack makes use of item metadata to reinforce the attack, such information is often publicly available, for example, the images/descriptions/titles on Amazon can be collected by directly browsing or writing crawlers.
In addition, since real-world RSs are usually complex and flexible (e.g., they may adopt ensemble schemes and may be updated frequently), it is difficult to gain accurate knowledge (e.g., model architectures, parameters and implementation details) of the victim RSs \cite{lin2022shilling}.
We therefore consider black-box settings for the victim RSs.

\subsubsection{\textbf{Attacker's Capability}} \label{attacker_budget}

To avoid being detected while conducting attacks under a budget, an attacker often injects a limited number of profiles \cite{lin2022shilling} and interacts with only a few items in each profile \cite{huang2021data}. As a result, an attacker's capabilities are restricted by the attack size (i.e., the number of injected fake user profiles $|\widetilde{\mathcal{U}}|$) and the profile size (i.e., the number of interactive items in each fake profile). To prevent suspicion while controlling the budget caused by frequent queries to the black-box target system, the attacker employs a surrogate agent to approximate its behavior. Consequently, our attack operates without relying on direct queries to the target system.

\subsection{Formulate Attacks As An Optimization Problem} \label{formulate_attacks}

\begin{figure*}[t]
    \centering
    \includegraphics[width=1\linewidth]{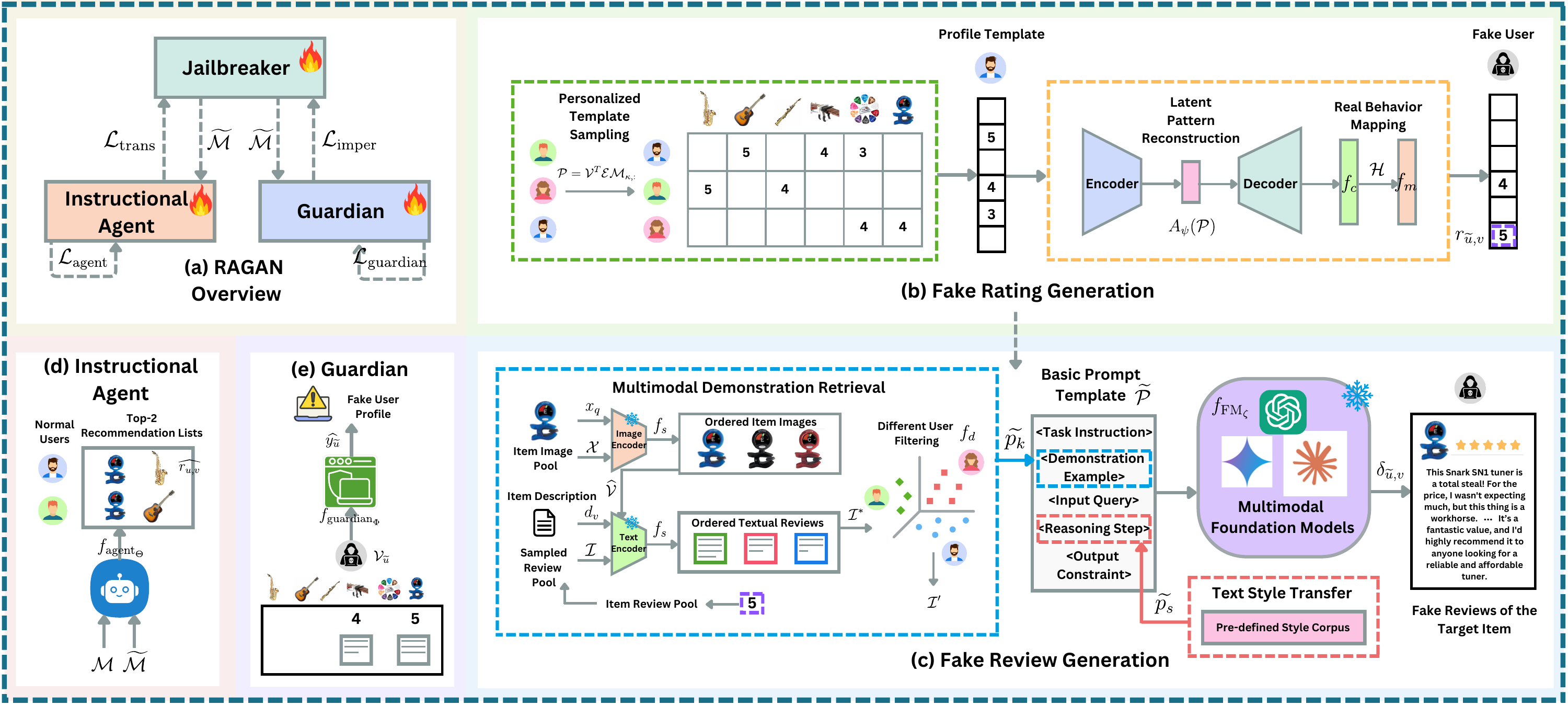}
    \vspace{-1.4em}
    \caption{The practical attack framework \textbf{RAGAN} consists of three modules: 1) a \textbf{Jailbreaker} that is responsible for generating fake user profiles containing both numerical ratings and textual reviews, where the quality of the profiles is strengthened through in-context learning enhanced with multimodal demonstration retrieval and text style transfer; 2) an \textbf{Instructional Agent} that is designed to improve attack transferability; and 3) a \textbf{Guardian} that is aimed at enhancing the imperceptibility of the profiles.}
    \label{fig:r_trojan++_overview}
    \vspace{-1.6em}
\end{figure*}

We formulate the attack as a bi-level optimization problem, as shown in Fig.~\ref{fig:r_trojan++_overview}(a), and solve for this problem to obtain high-quality fake user profiles, as inspired by \cite{lin2022shilling, yang2023incorporated}. Our attack consists of a jailbreaker, an instructional agent and a guardian. \textit{The lower-level} computes the optimal parameters of the agent (denoted by $\Theta$) and the guardian (denoted by $\Phi$) with the given benign matrix $p\mathcal{M}$ and fake matrix $\widetilde{\mathcal{M}}$. \textit{The upper-level} optimizes $\widetilde{\mathcal{M}}$ to maximize the attack objectives based on model parameters obtained by solving the lower-level problem.
\begin{equation} \label{eq_1}
\begin{aligned}
\mathop{\min}_{\widetilde{\mathcal{M}}}\ \lambda  \mathcal{L}_{\text{trans}}(\widehat{\mathcal{M}_{\Theta}}) + (1-\lambda) \mathcal{L}_{\text{imper}}(\widehat{\widetilde{\mathcal{M}}_{\Phi}}) \\
\text{subject to}\ \Theta = \mathop{\arg\min}_{\Theta} \mathcal{L}_{\text{agent}}(\mathcal{M}^*, \widehat{\mathcal{M}_{\Theta}^*})\\
\text{and}\ \Phi = \mathop{\arg\max}_{\Phi} \mathcal{L}_{\text{guardian}}(\mathcal{M}^*, \widehat{\mathcal{M}_{\Phi}^*}),
\end{aligned}
\end{equation} where $\mathcal{M}^*=\text{concatenation}(p\mathcal{M};\widetilde{\mathcal{M}})$, and  $\widehat{\mathcal{M}_{\Theta}^*}$ and $\widehat{\mathcal{M}_{\Phi}^*}$ are predictions from the corresponding models with parameters $\Theta$ and $\Phi$, respectively. $\mathcal{L}_{\text{agent}}$ and $\mathcal{L}_{\text{guardian}}$ denote the training objectives of the related modules. $\mathcal{L}_{\text{trans}}$ is the transferability objective defined on normal user's predictions $\widehat{\mathcal{M}_{\Theta}}$, $\mathcal{L}_{\text{imper}}$ is the imperceptibility objective defined on fake user's predictions $\widehat{\widetilde{\mathcal{M}}_{\Phi}}$ and $\lambda$ is a configurable parameter to adjust the trade-off of two objectives when optimized simultaneously. For simplicity, we use $\mathcal{M}$ to represent the partially observed matrix $p\mathcal{M}$ in the following formulations.

\section{RAGAN}
In this section, we present the practical attack framework RAGAN for high-quality fake user profile generation. RAGAN enhances the quality (i.e., transferablity and imperceptibility) of the profiles within the limited accessible resources by introducing publicly available textual reviews and further reinforce the profiles by improving the quality of the textual reviews. Fig.~\ref{fig:r_trojan++_overview} shows an overview of RAGAN. RAGAN is a tailored GAN that is composed of a jailbreaker, an instructional agent and a guardian, as shown in the figure. The design of each module is elaborated as follows.

\subsection{Jailbreaker}
Jailbreaker is a novel attack generator that aims to produce transferable and imperceptible fake user profiles. Unlike previous generators that manipulate ratings or reviews in isolation, Jailbreaker jointly models users’ numerical and textual behaviors to ensure realism. Given that textual reviews are increasingly incorporated to improve the RSs \cite{xu2021understanding}, and that informative reviews play an important role in explaining the reasons for the numerical ratings convincingly \cite{yang2023incorporated} and influencing users' decisions \cite{mohawesh2021fake}, each attack profile is created with synthetic numerical ratings and enhanced with corresponding fabricated reviews. To this end, we introduce a behavior pattern learning scheme to produce fake numerical ratings. For the generation of the high-quality fake textual reviews, we propose a demonstration retrieval algorithm and a text style transfer strategy to harness the in-context learning capabilities of the multimodal FMs.

\subsubsection{\textbf{Fake Numerical Rating Generation}}
The behavior pattern learning scheme is composed of three critical components to ensure the quality of the rating parts of the profiles, as shown in Fig.~\ref{fig:r_trojan++_overview}(b).

\textbf{\textit{Personalized Template Sampling.}} Some works  (e.g., \cite{zhang2021attacking}) create attack profiles from scratch (i.e., noise), resulting in low-quality profiles being produced. To address this, inspired by \cite{lin2022shilling, lin2020attacking}, we use benign user profiles containing real rating patterns as templates. Unlike these previous methods that randomly sample from $\mathcal{M}$, we prioritize user profiles based on three standards: 1) preference is given to profiles that have interacted with items similar to the target item (e.g., within the same category), as they better reflect realistic behaviors given that benign users typically do not interact with items arbitrarily (e.g., guitarists tend not to purchase drum-related products \cite{yang2023incorporated}) and are more susceptible to influence due to the fact that users share similar tastes; 2) profiles exhibiting a rich interaction history are favored to enhance personalization; and 3) profiles containing the target item are explicitly excluded to align with target normal users. As such, these standards collectively improve the effectiveness and imperceptibility of the attack. Let $\mathcal{P}$ to denote the profile templates,
\begin{equation} \label{eq_2}
\mathcal{P} = (\mathcal{E}\mathcal{M}_{r_{\widehat{\mathcal{U}}_t,:}})_{0:|\widetilde{\mathcal{U}}|,:}, 
\end{equation} where the rating part of $\mathcal{M}$, represented as $\mathcal{M}_r$, $\mathcal{M}_r \in \mathbb{R}^{|\mathcal{U}| \times |\mathcal{V}|}$. $\widehat{\mathcal{U}}_t$ indicates the normal user set to align with 3), and $\mathcal{E}$ denotes a permutation matrix for sorting the matrix to match 1) and 2), where $\mathcal{E} \in \mathbb{R}^{|\widehat{\mathcal{U}}_t| \times |\widehat{\mathcal{U}}_t|}$.

\textbf{\textit{Latent Pattern Reconstruction.}} To capture complex user-item associations from the templates in the latent space, RAGAN uses neural network paradigms with non-linearity \cite{wu2020densely, yang2021hunter, yang2021dualnet, yang2021deep} for the pattern reconstruction. Formally, the architecture of the paradigm is \begin{gather} \label{eq_3}
A_{\psi}(\mathcal{P}) = f_{l}(\cdots f_2(f_1(\mathcal{P})))\cdots),
\end{gather} where $\mathcal{A}$ denotes the paradigm. There are various possible implementations for the paradigm. Here, we adopt AutoEncoder \cite{zhang2019deep} (with parameters $\psi$) to sufficiently learn features at various levels of abstraction. AutoEncoder is built with an encoder and a decoder, where the encoder extracts the feature representations (i.e., user preferences) from the input and the decoder regenerates the rating vectors based on the representations. We use multi-layer perceptrons (MLPs) to build the function $f_j(\cdot), \ j=1, 2, ..., l$, where $f_j(x)=\text{Dropout}(\text{ReLU}(\text{BatchNorm}(W_jx+b_j)))$, and $W_j$ and $b_j$ are learnable parameters. For the encoder, we set the dimension of layers to half the dimension of previous layers; and for the decoder, which is a symmetric structure of an encoder, i.e., the size of layers is set to twice the size of the previous layers.

To align the output of the AutoEncoder with the real rating approach in RSs (e.g., Likert-scale \cite{yang2023incorporated} that is the most commonly used), RAGAN rescales the output  within a certain range (e.g., [1, 5]), as formulated below.
\begin{equation} \label{eq_4}
f_{c}(\mathcal{A}_{\psi}(\mathcal{P})) = \frac{r_{u,v}^{\text{max}} - r_{u,v}^{\text{min}}}{1+e^{-(W_{c}\mathcal{A}_{\psi}(\mathcal{P})+b_{c})}} + r_{u,v}^{\text{min}}
\end{equation} where $f_c(\cdot)$ is the reconstruction function, $r_{u,v}^{\text{max}}$ is the maximum rating of $\mathcal{M}$ (e.g., 5), $r_{u,v}^{\text{min}}$ is the minimum rating of $\mathcal{M}$ (e.g., 1) and $W_{c}$ as well as $b_{c}$ are learnable parameters.

\begin{figure*}[t]
    \centering
    \includegraphics[width=\linewidth]{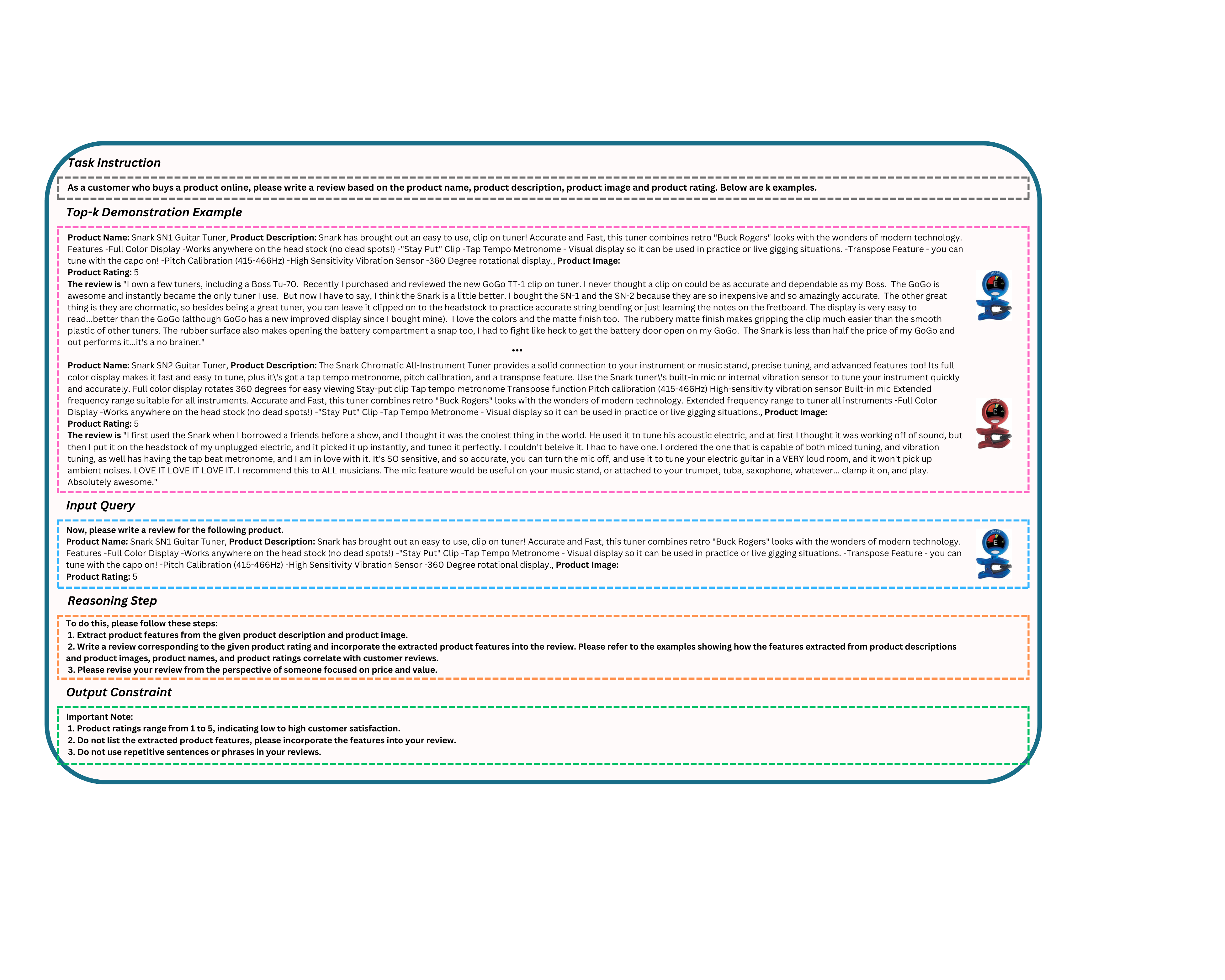}
    \caption{An example of a multimodal prompt within the well-crafted prompt template that is fed into a foundation model for fake review generation, where top-$k$ demonstration examples are obtained by our multimodal demonstration retrieval algorithm and our text style transfer strategy is integrated into the chain-of-thought reasoning process.}
    \label{fig:prompt_template}
    \vspace{-1.5em}
\end{figure*}

\textbf{\textit{Real Behavior Mapping.}} As discussed in previous sections, RAGAN aims to maximize $h(t)$. Thus, for each attack profile, the rating of the target item is increased to the maximum rating of $\mathcal{M}$, i.e., $r_{u,v}^{\text{max}}$, so as to promote the target item to as many normal users as possible to improve the attack effectiveness. For the filler items, an rounding-off operation $f_o(\cdot)$ is first used to discretize their ratings to match the input characteristics of the RSs. Given that the recommendation space in the real world is very large, each benign user generally interacts with only a few items in the space \cite{yang2023incorporated}, such as a customer cannot purchase all the products on Amazon. Hence, to align with the real behaviors to avoid being discovery, RAGAN then masks most harmful and useless ratings of the generated profiles via a masking matrix $\widetilde{\mathcal{E}}$. Formally, the overall operation  $f_m(\cdot)$ is
\begin{equation} \label{eq_5}
 f_m(\mathcal{H}) = \widetilde{\mathcal{E}} \circ f_o(\mathcal{H}),
 \end{equation} where $\mathcal{H}=f_c(\mathcal{A}_{\psi}(\mathcal{P})) \in \mathbb{R}^{|\widetilde{\mathcal{U}}|\times|\mathcal{V}|}$. $\widetilde{\mathcal{E}} \in \mathbb{R}^{|\widetilde{\mathcal{U}}|\times|\mathcal{V}|}$ performs Hadamard product, where each entry is $1$ or $0$ representing the filler item to be retained or not.
 Let $\mathcal{P}=(p_{\widetilde{u},v})_{|\widetilde{\mathcal{U}}|\times|\mathcal{V}|}$, $\widetilde{\mathcal{E}}=(\widetilde{e}_{\widetilde{u},v})_{|\widetilde{\mathcal{U}}|\times|\mathcal{V}|}$, $\mathcal{H}=(h_{\widetilde{u},v})_{|\widetilde{\mathcal{U}}|\times|\mathcal{V}|}$, and $\mathcal{F}$ to denote the profile size. For each malicious user $\widetilde{u}$'s profile, when the absolute difference between $h_{\widetilde{u},v}$ and $p_{\widetilde{u},v}$ belongs to the first $\mathcal{F}-1$ smallest differences and $p_{\widetilde{u},v} \neq 0$, $\widetilde{e}_{\widetilde{u},v} = 1$; otherwise, $\widetilde{e}_{\widetilde{u},v} = 0$. Thus, the behavior patterns of real users can be preserved to the maximum extent, which further improves diversity (e.g., picky users vs. nicey users) and thus the profiles' stealthiness.

\subsubsection{\textbf{Fake Textual Review Generation}} Multimodal FMs have shown remarkable capabilities in language and vision understanding, reasoning
and generation, revolutionizing a wide range of fields \cite{huang2024securing}. In-context learning empowers FMs to generalize to unseen downstream tasks purely through prompt-based conditioning, eliminating the need for parameter updates. Compared to fine-tuning and re-training approaches, which may be impractical due to restricted access to the model parameters (e.g., closed source models like GPT-4 and Claude 3) or constraints on computational resources required for large models such as LLaMA 3.3, such an effective and efficient paradigm gains increasing popularity \cite{luo2024context}.  As such, we carefully craft prompts to effectively harness the ICL capabilities of the FMs to improve the quality of the corresponding fake textual reviews, as shown in Fig.~\ref{fig:r_trojan++_overview}(c). Enhancing review quality can strengthen the optimization procedure of the attack effectiveness, resulting in more effective attack profiles.

\textbf{\textit{Basic Prompt Template.}} We prepare the prompt template $\widetilde{\mathcal{P}}$ to provide instructions, goals, demos, queries, constraints and other contexts, so that the model knows `what needs to be done' and `how to respond to', as shown in Fig.~\ref{fig:prompt_template}. The main components are elaborated below.

\begin{enumerate}
    \item \textit{Task Instruction $\widetilde{p_t}$.}  To make the generated reviews adapt to specific scenarios, we introduce role information \cite{yang2025drunkagent} into context, allowing the FMs to respond with domain-specific knowledge. We then give detailed instructions, which guides the model to understand what tasks need to be done and generate desired responses.
    
    \item \textit{Demonstration Example $\widetilde{p_d}$.} Given that ICL often learns the pattern hidden in the demonstrations and accordingly makes predictions \cite{luo2024context}, the demonstrations play an important role in outputting expectantly.  As such, we adopt few-shot ICL. As the figure shows, to sufficiently take advantage of the \textit{internal knowledge} of the FMs,  product names are introduced into the demos. To enrich the demos to improve the quality of the reviews, \textit{external resources} across multiple modalities (e.g., product descriptions and images) are incorporated. For aligning the review with the given rating, product ratings is added.

    \item \textit{Input Query $\widetilde{p_i}$.} We present the task-specific queries corresponding to the items and ratings of the obtained fake user profiles. As shown in the example, we maintain the same product attributes and presentation order as demonstrations to fully leverage the language and vision understanding and reasoning capabilities of the FMs to improve the effectiveness of the generation.

    \item \textit{Reasoning Step $\widetilde{p_r}$.} To enrich user and item features in the generated reviews and enhance the alignment between the reviews and the ratings to ensure the quality, we adopt chain-of-thought (CoT) \cite{wei2022chain} to instruct the model how to respond to precisely.

    \item \textit{Output Constraint $\widetilde{p_o}$.} We give some constraints on the response to facilitate subsequent optimizations. Specifically, we define a rating range to further strengthen the alignment and impose additional format limits informed by empirical observations to reduce undesirable model behaviors that undermine review qualities.

\end{enumerate}

Hence, the adversarial prompt template can be formulated:
\begin{gather} \label{eq_6}
\widetilde{\mathcal{P}} = \widetilde{p_t} \oplus \widetilde{p_d} \oplus \widetilde{p_i} \oplus \widetilde{p_r} \oplus \widetilde{p_o},
\end{gather} where $\oplus$ denotes the
integration of textual strings.

Based on our observations, when demonstrations are overly generic, the generated reviews tend to lack specificity in item characteristics and diversity in user behaviors, for instance, a review like ``The Behringer A500 is a great amplifier for the price. It is powerful and has a lot of features.''. In reality, users focus on different aspects of an item such as ease of use, affordability, flexibility, or scalability, based on their individual preferences. Moreover, users exhibit varied personalities (e.g., lenient or picky), thinking patterns, and cultural backgrounds, all of which shape their reviewing styles. The \textit{richer the item features} described in textual reviews, the more users with similar tastes can be identified, thereby \textit{increasing the promotional effect}. In addition, \textit{the greater the variation in item features across different reviews of the target item} and \textit{the more divergent the user features across profiles}, the \textit{less detectable} the attack becomes \cite{lin2022shilling}. Hence, the general reviews largely limit the effectiveness and stealthiness of the attack. To address the limitations, we propose a multimodal demonstration retrieval algorithm and a text style transfer strategy to enrich review quality and disguise.

\textbf{\textit{Multimodal Demonstration Retrieval.}} We present an algorithm to retrieve review texts containing rich semantics and diverse behavior patterns from the learning corpus as top-k demonstrations. Given that the ICL performance is sensitive not only to the selection of demo samples, but also to the order of demonstrations \cite{luo2024context}, the algorithm prioritizes the reviews that are the most relevant to the query to enhance the generation quality of reviews. 

\begin{figure}[t]
    \centering
    \includegraphics[width=0.8\linewidth]{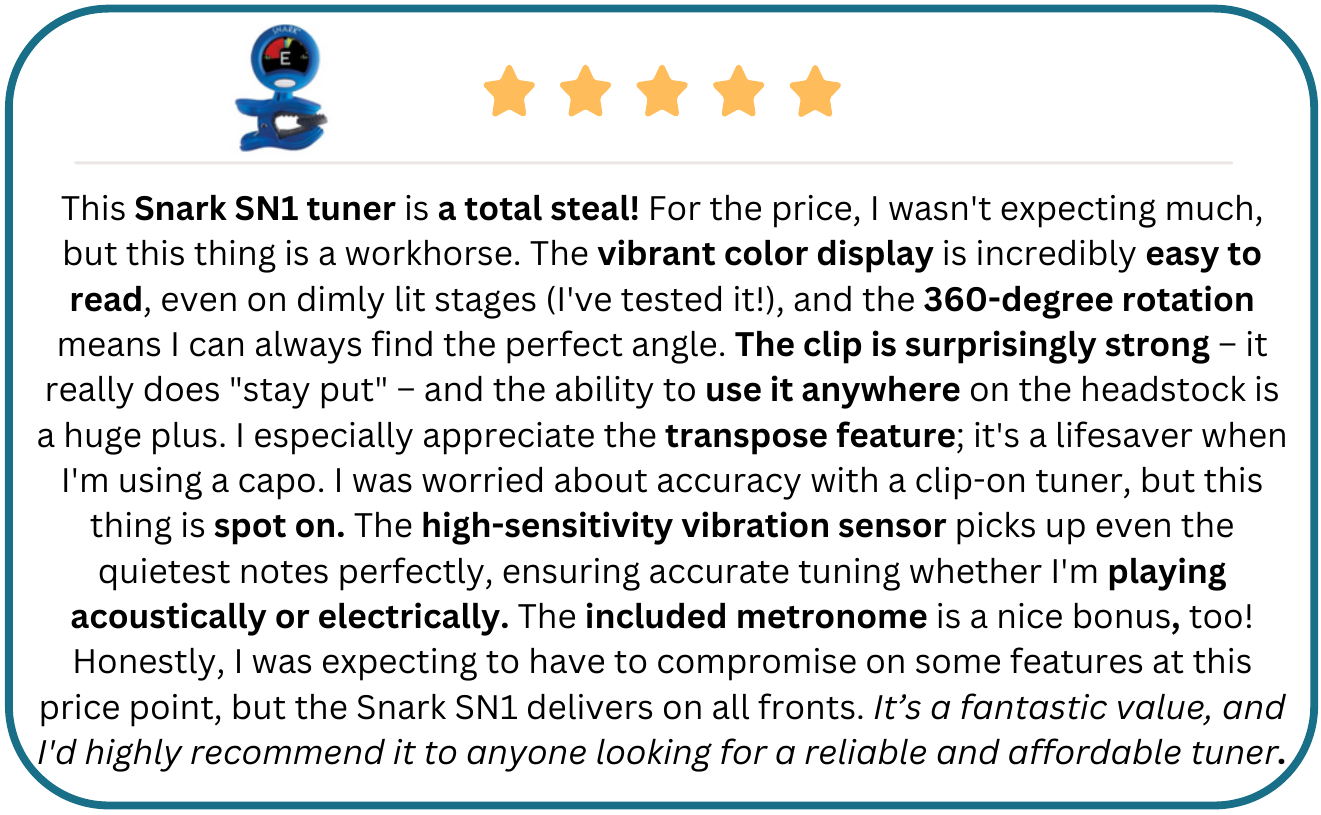}
    \caption{An example of reviews for a target item in the fake user profiles generated by RAGAN.}
    \label{fig:gen_reviews}
    \vspace{-1.5em}
\end{figure}

For each rated item of each fake user $\widetilde{u}$'s profile (i.e., the target item and the filler items marked by Eq.~(\ref{eq_5})), $k$ appropriate and related multimodal demonstrations are retrieved, as shown in Fig.~\ref{fig:r_trojan++_overview}(c).
First, the algorithm obtains a set of items that are most similar to the rated item. Given that the visual appearances of items provide rich feature elements and crucial signals about user preferences \cite{yang2024attacking}, we propose to use the item images to calculate the similarity. For each item image $x_{v}, x_{v} \in \mathcal{D}$, the similarity score between it and the query image $x_q$ (i.e., the rated item's image) is obtained by a function $f_s(\cdot)$,
\begin{equation} \label{eq_7}
s_v = f_s(x_q,x_v),
\end{equation} where $s_v$ represents the similarity score and $f_s(\cdot)$ is implemented via lightweight cosine similarity, i.e., $f_s(x_q,x_v) = \frac{f_e(x_q)\cdot f_e(x_v)}{||f_e(x_q)||||f_e(x_v)||}$. $f_e(\cdot)$ refers to a pre-trained vision encoder for feature vector extraction. We use CLIP-ViT\footnote{https://openai.com/index/clip/} to build it here. As a result, an ordered list of items $\widehat{\mathcal{V}}$ can be produced using the sequence of the similarity scores $\{s_v\}$. Formally, 
$\widehat{\mathcal{V}} = \mathop{\text{argsort}}_{v}(\{s_v\})$,  where the higher the score, the higher the ranking of the item. Note that the first one of $\widehat{\mathcal{V}}$ is the rated item itself.

To ensure the alignment between the numerical ratings and the textual reviews, the algorithm then selects the reviews from the matrix $\mathcal{M}$ with the same rating of the attack item (i.e., $r_{\widetilde{u},v}$). As such, such demos have high similarity to the input query $\widetilde{p_i}$. Formally, $\mathcal{I} = \{\delta_{u,v}: r_{u,v} = r_{\widetilde{u},v}, r_{u,v} \in \mathcal{M} \}$, where $\mathcal{I}$ is the set of the filtered reviews.

Afterwards, the algorithm sorts the selected reviews in $\mathcal{I}$ based on the obtained similar items and the semantic richness. The algorithm measures the similarity between the reviews and the textual item descriptions $d_v, d_v \in \mathcal{D}$ (including material, volume, shape, weight, colour, functionality, usage scenarios and so on) to get the semantic richness scores. For each user $u, u \in \mathcal{U}$'s textual review $\delta_{u,v}$, $\delta_{u,v} \in \mathcal{I}$, the richness score $c_{u,v}$ is computed through
\begin{equation} \label{eq_10}
c_{u,v} = f_s(d_v,\delta_{u,v}),
\end{equation} 
where $f_s(\cdot)$ is implemented via the cosine similarity. Thereinto, $f_e(\cdot)$ is constructed with a pre-trained textual encoder, sentence transformer (a.k.a. SBERT\footnote{https://sbert.net/}). If $d_v$ is empty, the algorithm uses the lexical diversity to indicate the richness, that is, the number of different words in the textual review. The algorithm thus ranks the reviews for each $v, v \in \widehat{\mathcal{V}}$ based on $c_{u,v}$. Let $\mathcal{I}_v$ denote the sequence of $v$'s ordered review samples and $\mathcal{I}^*$ represent the list of all the reviews,
\begin{equation} \label{eq_11}
\begin{aligned}
\mathcal{I}_v = \{ \delta_{u,v} \}\\
\text{subject to}\ u = \mathop{\text{argsort}}_{u}(\{c_{u,v}\}), 
\end{aligned}
\end{equation} where the higher the score, the richer the item features of the review text. Hence, $\mathcal{I}^*=\text{concatenation}(\{\mathcal{I}_v\})$.

\begin{algorithm}[!t]
\caption{RAGAN Optimization Procedure}
\label{alg:training_procedure}

\renewcommand{\algorithmicrequire}{\textbf{Input:}}

\renewcommand{\algorithmicensure}{\textbf{Output:}}

\begin{algorithmic}[1]
\REQUIRE $p$ percent of user-item interaction matrix $\mathcal{M}$, and item metadata $\mathcal{D}$  
\ENSURE fake matrix $\widetilde{\mathcal{M}}$    

\FOR{number of training epochs}

\STATE \textbf{Solve for the Lower-level Problems:}

\STATE Obtain $\widetilde{\mathcal{M}}$ with Eq.~(\ref{eq_2}) - Eq.~(\ref{eq_13})

\FOR{number of training epochs}
\STATE Optimize the Instructional Agent $\Theta$ on $p\mathcal{M}+\widetilde{\mathcal{M}}$  to minimize $\mathcal{L}_{\text{agent}}$
\ENDFOR

\STATE Optimize the Guardian $\Phi$ on $p\mathcal{M}+\widetilde{\mathcal{M}}$  to maximize $\mathcal{L}_{\text{guardian}}$

\STATE \textbf{Solve for the Upper-level Problem:}

\STATE Calculate $\mathcal{L}_{\text{trans}}$ on $\Theta$ with Eq.~(\ref{eq_14}) and Eq.~(\ref{eq_16})

\STATE Calculate $\mathcal{L}_{\text{imper}}$ on $\Phi$ with Eq.~(\ref{eq_15}) and Eq.~(\ref{eq_17})

\STATE Optimize the Jailbreaker to minimize Eq.~(\ref{eq_1})

\ENDFOR

\STATE Obtain optimal $\widetilde{\mathcal{M}}$ from the optimal Jailbreaker

\RETURN $\widetilde{\mathcal{M}}$

\end{algorithmic}

\end{algorithm}

The algorithm finally filters out the reviews from the same users and then obtains top-$k$ ordered reviews. Thus, a new review list $\mathcal{I}'$ for demo examples is created,
\begin{equation} \label{eq_12}
\mathcal{I}' = f_d(\mathcal{I}^*)_{:k},
\end{equation} where $f_d(\cdot)$ denotes the function to capture the reviews from different users for demonstrating diverse user features.

Let $\widetilde{p}_k$ denote the set of the multimodal demos. Based on our designed template, $\widetilde{p}_k=\{ (x_v, n_v, d_v, r_{u,v}, \delta_{u,v}): \delta_{u,v} \in \mathcal{I}', r_{u,v} \in \mathcal{M}, x_v, n_v, d_v \in \mathcal{D} \}$, which serves as in-context demo examples for FMs, aiming to enhance their understanding and performance on analogous tasks (i.e., review generation here).

\textit{\textbf{Text Style Transfer.}} Text style transfer is the task of rewriting text to incorporate additional stylistic elements while preserving the overall semantics and structure \cite{reif2021recipe}. To further enrich the user features, we adopt a strategy to transfer the style of the reviews. The strategy first establishes a comprehensive text style transfer prompt corpus encompassing ten distinct categories, as elaborated in the appendix. Then, the strategy randomly selects a candidate from the corpus as $\widetilde{p_s}$, such as `Please revise your review from the perspective of someone focused on price and value.'. The strategy finally incorporates $\widetilde{p_s}$ into the reasoning steps of the prompt template $\widetilde{\mathcal{P}}$, as shown in Fig.~\ref{fig:prompt_template}.

Consequently, for each rating $r_{\widetilde{u},v}$ obtained from $f_m(\mathcal{H})$, the corresponding review is obtained via
\begin{gather} \label{eq_13}
\delta_{\widetilde{u},v} = f_{\text{FM}_{\zeta}}(\widetilde{\mathcal{P}} \oplus \widetilde{p_k} \oplus \widetilde{p_s}),
\end{gather} where FM is with frozen parameters $\zeta$. As can be seen from the example in Fig. \ref{fig:gen_reviews}, the generated textual reviews contain rich item characteristics (bond fonts, e.g., excellent design and versatile functionality) and diverse user behaviors (italic fonts, e.g., cost-effective).

\subsection{Instructional Agent}

It is difficult to obtain accurate knowledge of the victim RSs, their detailed recommendation results, and periodic feedback in real-world scenarios \cite{yang2023incorporated}. Moreover, frequent queries to the target system can raise suspicion and often require a large attack budget. As such, we introduce an instructional agent to measure how effective the attack is and then optimize the Jailbreaker to improve the attack transferability and stealth while reducing the budget under black-box settings, as inspired by \cite{lin2022shilling, yang2023incorporated}. Existing works \cite{lin2022shilling, wu2021triple, zhang2021data, tang2020revisiting} often use rating-only-based RSs to evaluate the effectiveness, which is unsuitable for our profiles containing both numerical ratings and textual reviews. We thus adopt the review-based RS \cite{sachdeva2020useful} to build the agent, as shown in Fig.~\ref{fig:r_trojan++_overview}(d), where the semantically rich textual reviews are introduced to enrich user and item features for accurately predicting user preferences, so that the valuable and effective instructions can be provided to Jailbreaker within the limited accessible data.

The agent performs a function $f_{\text{agent}_{\Theta}}(\cdot)$ with the trainable parameters $\Theta$ to predict the preference of the user $u$ on $v$, which is denoted as $\widehat{r_{u,v}}$ and can be formulated below.
\begin{equation} \label{eq_14}
\widehat{r_{u,v}} = f_{\text{agent}_{\Theta}}(\delta_{u},\delta_{v},q_u,q_v),
\end{equation} where $\delta_u$ and $\delta_v$ are, respectively, the set of historical reviews from $u$ and on $v$, and $q_u$ and $q_v$ indicate the user ID embeddings and the item ID embeddings, separately. $f_{\text{agent}_{\Theta}}(\cdot)$ is generally constructed with feature modeling and preference prediction components \cite{ sachdeva2020useful}. Following the surrogate architecture of the previous work \cite{yang2023incorporated}, a TextCNN \cite{xu2021understanding} is first used to extract features from the user and the item reviews. The features are then combined with the embeddings. Afterwards, the latent representations of the user network and the item network are concatenated and mapped to a shared feature space. Finally, a MLP followed by a Sigmoid is adopted for effectively conducting implicit recommendations.

\begin{table}
  \caption{Statistics of Datasets}
  \vspace{-1.2em}
  \begin{center}
  \small
  \label{tab:statistics}
  \resizebox{\columnwidth}{!}{
 \begin{tabular}{c|ccc|c}
\hline
  Dataset & \#Users & \#Items & \#Reviews & Sparsity \\
\hline
  Amazon Musical Instruments & 1,429 & 900 & 10,261 & 99.20\% \\ 
 Amazon Automotive & 2,928 & 1,835 & 20,473 & 99.62\% \\ 
  Yelp & 1,599 & 1,318 & 30,120 & 98.57\%\\ 
\hline
  \end{tabular}}
  \end{center}
  \vspace{-2em}
\end{table}

\begin{table*}
\caption{HR@10 and NDCG@10 of different attacks against various victim RSs on real-world datasets. We use bold fonts and asterisk symbols to denote the best performance and second best performance methods, respectively.
}
\vspace{-1.2em}
  \begin{center}
  \label{tab:comp_results}
  \resizebox{2\columnwidth}{!}{
  \begin{tabular}{c|c|c|c|c|c|c|c|c|c|c|c|c|c}
    \hline
     \multirow{2}{*}{Victim RS}  & \multirow{2}{*}{Dataset} & \multirow{2}{*}{Metric} & \multicolumn{11}{c}{Attack Method}\\

     \cline{4-14}
     
     & & &  {Random} & {Bandwagon}& {PGA}& {DCGAN} & {AUSH} & {DLA} & {RecUP} &  {Leg-UP}& {TrialAttack} &{R-Trojan} &{RAGAN}\\

    \hline
    \hline
    
     \multirow{6}{*}{WRMF} & \multirow{2}{*}{Musical} & HR  & 0.2649 & 0.2968 & 0.3904 & 0.3599 & 0.3781 & 0.4383 &  0.4020 & 0.4681 &  0.4804 & 0.5530* & \textbf{0.6313}\\
     
     & &  NDCG  & 0.1082 & 0.1211 & 0.1822 &  0.1957 & 0.1845 & 0.2208 &  0.1970 & 0.2374 & 0.2418 & 0.3617* & \textbf{0.4254}\\

     \cline{2-14}

     & \multirow{2}{*}{Automotive} & \multicolumn{1}{c|}{HR} & 0.1137 & \multicolumn{1}{c|}{0.1348} & 0.1563 & \multicolumn{1}{c|}{0.1695} & 0.1698 & \multicolumn{1}{c|}{0.2094} & 0.1730 & \multicolumn{1}{c|}{0.2305} & 0.2284  & 0.2693* & \textbf{0.3185} \\
     
      & & \multicolumn{1}{c|}{NDCG} & 0.0451 & \multicolumn{1}{c|}{0.0533} & 0.0772 & \multicolumn{1}{c|}{0.0818} & 0.0796 & \multicolumn{1}{c|}{0.0981} & 0.0781 & \multicolumn{1}{c|}{0.1232} & 0.1145 & 0.1677* & \textbf{0.2262}\\

     \cline{2-14}

     & \multirow{2}{*}{Yelp} & \multicolumn{1}{c|}{HR} & \multicolumn{1}{c|}{0.0939} & 0.0861 & \multicolumn{1}{c|}{0.1082} & 0.1017 & \multicolumn{1}{c|}{0.1231} & 0.1276 & \multicolumn{1}{c|}{0.1399} & 0.1360 & \multicolumn{1}{c|}{0.1464} & 0.1852* & \textbf{0.2494} \\ 
     
      & & \multicolumn{1}{c|}{NDCG} &  0.0635 & \multicolumn{1}{c|}{0.0563} & 0.0801 & \multicolumn{1}{c|}{0.0731} & 0.0799 & \multicolumn{1}{c|}{0.0893} & 0.0969  & \multicolumn{1}{c|}{0.1087} & 0.0998 & 0.1496* & \textbf{0.2011} \\

    \hline

      \multirow{6}{*}{NCF} & \multirow{2}{*}{Musical} & \multicolumn{1}{c|}{HR} & 0.1495 & \multicolumn{1}{c|}{0.1509} & 0.1633 & \multicolumn{1}{c|}{0.1807} & 0.2068 & \multicolumn{1}{c|}{0.2525} & 0.2104 & \multicolumn{1}{c|}{0.2438} & 0.2714 & 0.3382* & \textbf{0.4209}\\

     & & \multicolumn{1}{c|}{NDCG}   & 0.0633 & \multicolumn{1}{c|}{0.0575} & 0.0677 & \multicolumn{1}{c|}{0.0816} & 0.0944 & \multicolumn{1}{c|}{0.1305} & 0.0802 & \multicolumn{1}{c|}{0.1112} & 0.1503 & 0.2009* & \textbf{0.2526}\\

     \cline{2-14}

     & \multirow{2}{*}{Automotive} & \multicolumn{1}{c|}{HR}  & 0.1043 & \multicolumn{1}{c|}{0.1102} & 0.1192 & \multicolumn{1}{c|}{0.1383} & 0.1650 & \multicolumn{1}{c|}{0.1962} & 0.1827 & \multicolumn{1}{c|}{0.2038} & 0.2187 & 0.2679* & \textbf{0.3095} \\ 
     
      & & \multicolumn{1}{c|}{NDCG} & 0.0474 & \multicolumn{1}{c|}{0.0515} & 0.0604 & \multicolumn{1}{c|}{0.0854} & 0.0971 & \multicolumn{1}{c|}{0.1009} & 0.0935 & \multicolumn{1}{c|}{0.1109} & 0.1353 & 0.1594* & \textbf{0.1925} \\

     \cline{2-14}

     & \multirow{2}{*}{Yelp} & \multicolumn{1}{c|}{HR}  & 0.0784 & \multicolumn{1}{c|}{0.0667} & 0.0848 & \multicolumn{1}{c|}{0.0965} &  0.0991 & \multicolumn{1}{c|}{0.1159} & 0.1082 & 0.1328 & 0.1250 & 0.1729* 
     & \textbf{0.2040}\\ 
     
      &  & \multicolumn{1}{c|}{NDCG}  & 0.0472 & \multicolumn{1}{c|}{0.0337} & 0.0561 & \multicolumn{1}{c|}{0.0603} & 0.0553 & \multicolumn{1}{c|}{0.0752} & 0.0598 & 0.0844 & 0.0851 & 0.1162* & \textbf{0.1469}\\

      \hline

  \multirow{6}{*}{LightGCN} & \multirow{2}{*}{Musical} & \multicolumn{1}{c|}{HR} & \multicolumn{1}{c|}{0.0602} & 0.0718 & \multicolumn{1}{c|}{0.0893} & 0.0972 & \multicolumn{1}{c|}{0.1110} & 0.1509 & \multicolumn{1}{c|}{0.1023} & 0.1437 & \multicolumn{1}{c|}{0.1357} & 0.1858* & \textbf{0.2569}\\

     & & \multicolumn{1}{c|}{NDCG}  & \multicolumn{1}{c|}{0.0260} & 0.0370 & \multicolumn{1}{c|}{0.0413} & 0.0466 & \multicolumn{1}{c|}{0.0556} & 0.0786 & \multicolumn{1}{c|}{0.0517} & 0.0698 & \multicolumn{1}{c|}{0.0732} & 0.0983* & \textbf{0.1331} \\

     \cline{2-14} 
     
     & \multirow{2}{*}{Automotive} & \multicolumn{1}{c|}{HR} & \multicolumn{1}{c|}{0.0340} & 0.0343 & \multicolumn{1}{c|}{0.0589} & 0.0433 & \multicolumn{1}{c|}{0.0634} & 0.1102 & \multicolumn{1}{c|}{0.0735} & 0.1088 & \multicolumn{1}{c|}{0.1009} & 0.1536* & \textbf{0.2149}\\ 
     
      & & \multicolumn{1}{c|}{NDCG} & \multicolumn{1}{c|}{0.0189} & 0.0174  & \multicolumn{1}{c|}{0.0359} & 0.0238 & \multicolumn{1}{c|}{0.0324} & 0.0599 & \multicolumn{1}{c|}{0.0387} & 0.0632 & \multicolumn{1}{c|}{0.0681} & 0.0846* & \textbf{0.1363}\\

     \cline{2-14}

     & \multirow{2}{*}{Yelp} & \multicolumn{1}{c|}{HR} & \multicolumn{1}{c|}{0.0136} & 0.0117 & \multicolumn{1}{c|}{0.0246} &  0.0317 & \multicolumn{1}{c|}{0.0415} & 0.0602 & \multicolumn{1}{c|}{0.0434} & 0.0544 & \multicolumn{1}{c|}{0.0512} & 0.0926* & \textbf{0.1308} \\ 
     
      & & \multicolumn{1}{c|}{NDCG} & \multicolumn{1}{c|}{0.0059} & 0.0061 & \multicolumn{1}{c|}{0.0141} & 0.0197 & \multicolumn{1}{c|}{0.0265} & 0.0262  & \multicolumn{1}{c|}{0.0252} &  0.0354 & \multicolumn{1}{c|}{0.0359} &  0.0560* & \textbf{0.0961} \\ 

      \hline
  \end{tabular}}
  \end{center}
  \vspace{-2.5em}
\end{table*}

\subsection{Guardian}

The guardian with the learnable parameters $\Phi$ plays a minimax game with the jailbreaker empowered by the agent, as inspired by the vanilla GAN \cite{goodfellow2020generative}. Therefore, the module distinguishes the fake user profiles from the benign user profiles as much as possible to enhance the imperceptibility of the attack, as shown in Fig.~\ref{fig:r_trojan++_overview}(e). Let $f_{\text{guardian}_\Phi}(\cdot)$ denote the function to identify malicious/benign behavior patterns from fake/benign user $\widetilde{u}$/$u$. In the case of the fake user profile $\mathcal{V}_{\widetilde{u}}$,
\begin{equation} \label{eq_15}
\widehat{y_{\widetilde{u}}} = f_{\text{guardian}_\Phi}(\mathcal{V}_{\widetilde{u}}),
\end{equation} where $\widehat{y_{\widetilde{u}}}$ denotes the prediction of $\widetilde{u}$'s profile. There are two kinds of predicted results: one is normal and another is abnormal. Based on the previous detector \cite{yang2023incorporated}, $f_{\text{guardian}_\Phi}(\cdot)$ can be built through a TextCNN, an Encoder and a MLP, where the TextCNN enhanced by the Encoder are used to capture features from the profiles, and the MLP followed by a Sigmoid is leveraged to perform detections based on the obtained feature representations. The benign user profile's prediction $\widehat{y_u}$ can be produced via the similar operations.

\subsection{Learning}

As discussed in Section \ref{formulate_attacks}, we obtain high-quality attack profiles by solving a bi-level optimization problem.

\textit{\textbf{The Lower-level Problems.}}  We adopt binary cross-entropy (BCE) \cite{he2017neural} for $\mathcal{L}_{\text{agent}}$ to adapt to the implicit top-$\mathcal{K}$ recommendation task.  $\Theta$ will be obtained when $\mathcal{L}_{\text{agent}}$ is minimum that is closest to the actual situation. Similarly, given that the guardian performs a binary classification task, inspired by  \cite{lin2022shilling}, we adopt negative BCE for $\mathcal{L}_{\text{guardian}}$. Since this module plays the minimax game with the jailbreaker, $\Phi$ will be obtained when $\mathcal{L}_{\text{guardian}}$ is maximum, ensuring that the guardian achieves its best ability to distinguish between benign and attack profiles.

\textit{\textbf{The Upper-level Problem.}} $\mathcal{L}_{\text{trans}}$ aims to maximize $h(t)$. If $t$ is in the recommendation lists of the normal users, it is not necessary to optimize much. But if not, it is to minimize the prediction rating gap between $t$ and the predicted items that are in the recommended lists, so that the target item can be promoted to as many normal users as possible. The loss is designed as follows.
\begin{equation} \label{eq_16}
\mathcal{L}_{\text{trans}}=\log(\sum\limits_{u \in \widehat{\mathcal{U}}_t} \sum\limits_{v \in \widehat{\mathcal{V}}_{u,:k}} (e^{\widehat{r_{u,v}}} - e^{\widehat{r_{u,t}}})+1),
\end{equation} where $\widehat{r_{u,t}}$ is the prediction of the target normal user $u$ on the target item $t$ from $f_{\text{agent}_{\Theta}}(\cdot)$ and $e^{(\cdot)}$ is used to amplify the rating gap and $\log(\cdot)$ is used to shrink the overall sum of the gap to a range to avoid this attack objective overly dominating the optimization direction. To make sure the loss is positive, we add 1 to the sum. $\mathcal{L}_{\text{imper}}$ is optimized by tricking the guardian that the fake profiles are from real users, formally,
\begin{equation} \label{eq_17}
    \mathcal{L}_{\text{imper}} = \frac{1}{\widetilde{\mathcal{U}}} \sum\limits_{\widetilde{u} \in \widetilde{\mathcal{U}}} \log(1-\widehat{y_{\widetilde{u}}}).
\end{equation}

$\mathcal{L}_{\text{imper}}$ is weighted by $\lambda$ and combined with $\mathcal{L}_{\text{trans}}$ to form the final loss for optimizing the parameters of the generation module, as detailed in Section \ref{formulate_attacks}.  Inspired by the works \cite{lin2022shilling, yang2023incorporated}, the optimization procedure of RAGAN is detailed in Algorithm \ref{alg:training_procedure}, where Adam is used as an optimizer.


\begin{table*}
  \caption{HR@10 and NDCG@10 of R-Trojan and RAGAN with different attack sizes against various review-based RSs on three real-world datasets, respectively.}
  \vspace{-1em}
  \begin{center}
  \small
  \label{tab:review_results}
  \resizebox{2\columnwidth}{!}{
  \begin{tabular}{c|c|cccccccccccc}
    \hline
     \multirow{3}{*}{Dataset} & \multirow{3}{*}{Attack} & \multicolumn{6}{c|}{DeepCoNN} & \multicolumn{6}{c}{Review-based Agent}\\

    \cline{3-14} & & \multicolumn{2}{c|}{0.5\%} & \multicolumn{2}{c|}{1\%} & \multicolumn{2}{c|}{3\%} & \multicolumn{2}{c|}{0.5\%} & \multicolumn{2}{c|}{1\%} & \multicolumn{2}{c}{3\%}  \\
    
     \cline{3-14} & & HR & \multicolumn{1}{c|}{NDCG} & HR & \multicolumn{1}{c|}{NDCG} & HR & \multicolumn{1}{c|}{NDCG}  &  HR & \multicolumn{1}{c|}{NDCG} & HR & \multicolumn{1}{c|}{NDCG} & HR & \multicolumn{1}{c}{NDCG}\\
    
    \hline
    \multirow{3}{*}{Musical} & \multicolumn{1}{c|}{None} & 0.0000 &\multicolumn{1}{c|}{0.0000} & 0.0000 & \multicolumn{1}{c|}{0.0000}& 0.0000 &\multicolumn{1}{c|}{0.0000} & 0.0000 &\multicolumn{1}{c|}{0.0000} & 0.0000 &\multicolumn{1}{c|}{0.0000}& 0.0000 &\multicolumn{1}{c}{0.0000}\\

   & R-Trojan & 0.1219 &\multicolumn{1}{c|}{0.0356} & 0.1916 & \multicolumn{1}{c|}{0.0560}& 0.3991 &\multicolumn{1}{c|}{0.1170} & 0.1938 &\multicolumn{1}{c|}{0.0583} & 0.2663 &\multicolumn{1}{c|}{0.0783}& 0.5835 &\multicolumn{1}{c}{0.2132} \\

   & RAGAN & \textbf{0.3316} &\multicolumn{1}{c|}{\textbf{0.0972}} & \textbf{0.4049}  & \multicolumn{1}{c|}{\textbf{0.1191}} & \textbf{0.5624} &\multicolumn{1}{c|}{\textbf{0.2018}} & \textbf{0.3752} &\multicolumn{1}{c|}{\textbf{0.1114}} & \textbf{0.4521} &\multicolumn{1}{c|}{\textbf{0.1353}} & \textbf{0.7438} &\multicolumn{1}{c}{\textbf{0.2334}}\\

    \hline
    \multirow{3}{*}{Automotive} & \multicolumn{1}{c|}{None} &  0.0000 &\multicolumn{1}{c|}{0.0000} & 0.0000 & \multicolumn{1}{c|}{0.0000} & 0.0000 &\multicolumn{1}{c|}{0.0000} & 0.0000 &\multicolumn{1}{c|}{0.0000} &  0.0000 &\multicolumn{1}{c|}{0.0000} & 0.0000 &\multicolumn{1}{c}{0.0000}\\

   & R-Trojan & 0.0810 &\multicolumn{1}{c|}{0.0287} & 0.1230  & \multicolumn{1}{c|}{0.0420} & 0.2763  &\multicolumn{1}{c|}{0.0939} &  0.1220 &\multicolumn{1}{c|}{0.0386} & 0.2026 &\multicolumn{1}{c|}{0.0823} & 0.4523 &\multicolumn{1}{c}{0.1576} \\

   & RAGAN & \textbf{0.1783} &\multicolumn{1}{c|}{\textbf{0.0522}} & \textbf{0.2192} & \multicolumn{1}{c|}{\textbf{0.0638}} & \textbf{0.4621} &\multicolumn{1}{c|}{\textbf{0.1963}} & \textbf{0.2293}  & \multicolumn{1}{c|}{\textbf{0.0704}} &  \textbf{0.3330} & \multicolumn{1}{c|}{\textbf{0.1275}} & \textbf{0.6312} &\multicolumn{1}{c}{\textbf{0.2245}} \\

 \hline
    \multirow{3}{*}{Yelp} & \multicolumn{1}{c|}{None} &  0.0000 & \multicolumn{1}{c|}{0.0000} & 0.0000 & \multicolumn{1}{c|}{0.0000} &0.0000  &\multicolumn{1}{c|}{0.0000} & 0.0000 &\multicolumn{1}{c|}{0.0000} & 0.0000 &\multicolumn{1}{c|}{0.0000} & 0.0000 &\multicolumn{1}{c}{0.0000}\\

   & R-Trojan & 0.0547 &\multicolumn{1}{c|}{0.0160} & 0.0861 & \multicolumn{1}{c|}{0.0252}& 0.2003 &\multicolumn{1}{c|}{0.0586} & 0.0835 &\multicolumn{1}{c|}{0.0252} & 0.1574 &\multicolumn{1}{c|}{0.0471} & 0.3297 &\multicolumn{1}{c}{0.1050}\\

   & RAGAN & \textbf{0.0835} &\multicolumn{1}{c|}{\textbf{0.0245}} & \textbf{0.1159} & \multicolumn{1}{c|}{\textbf{0.0341}} & \textbf{0.2597} &\multicolumn{1}{c|}{\textbf{0.0760}} & \textbf{0.1645} &\multicolumn{1}{c|}{\textbf{0.0504}} & \textbf{0.2545} &\multicolumn{1}{c|}{\textbf{0.0788}} & \textbf{0.4495} & \multicolumn{1}{c}{\textbf{0.1394}} \\

\hline
  
  \end{tabular}}
  \end{center}
\vspace{-1.5em}
\end{table*}

\begin{figure*}
\centering
\begin{subfigure}{.24\textwidth}
    \centering
    \includegraphics[width=\linewidth]{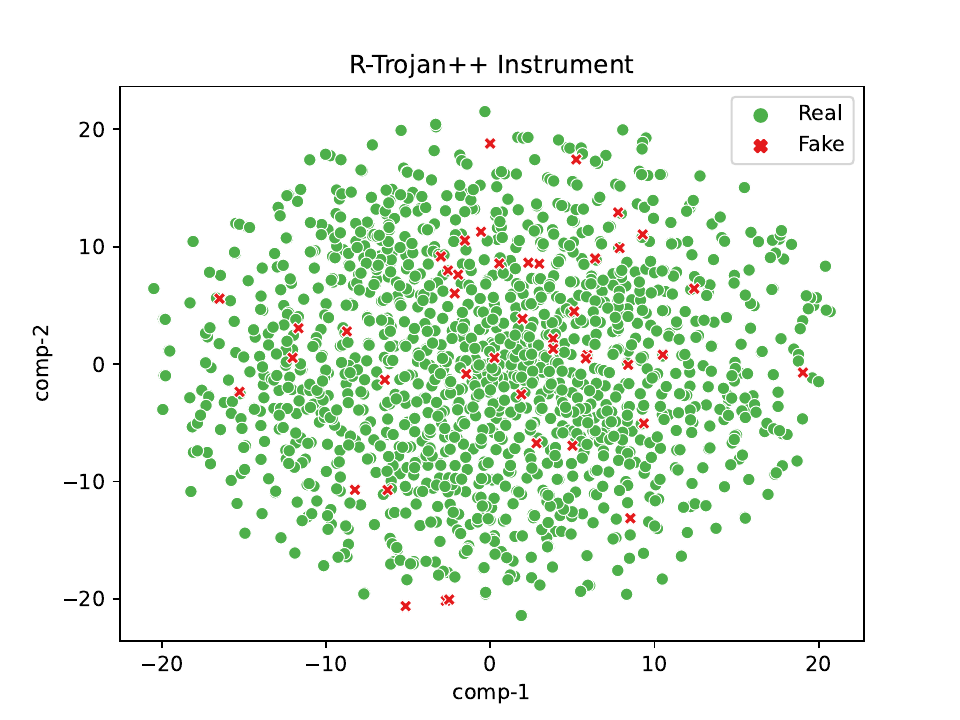}
    \vspace{-1.2em}
    \caption{RAGAN on Musical}
    \label{SUBFIGURE LABEL 1}
\end{subfigure}
\begin{subfigure}{.24\textwidth}
    \centering
    \includegraphics[width=\linewidth]{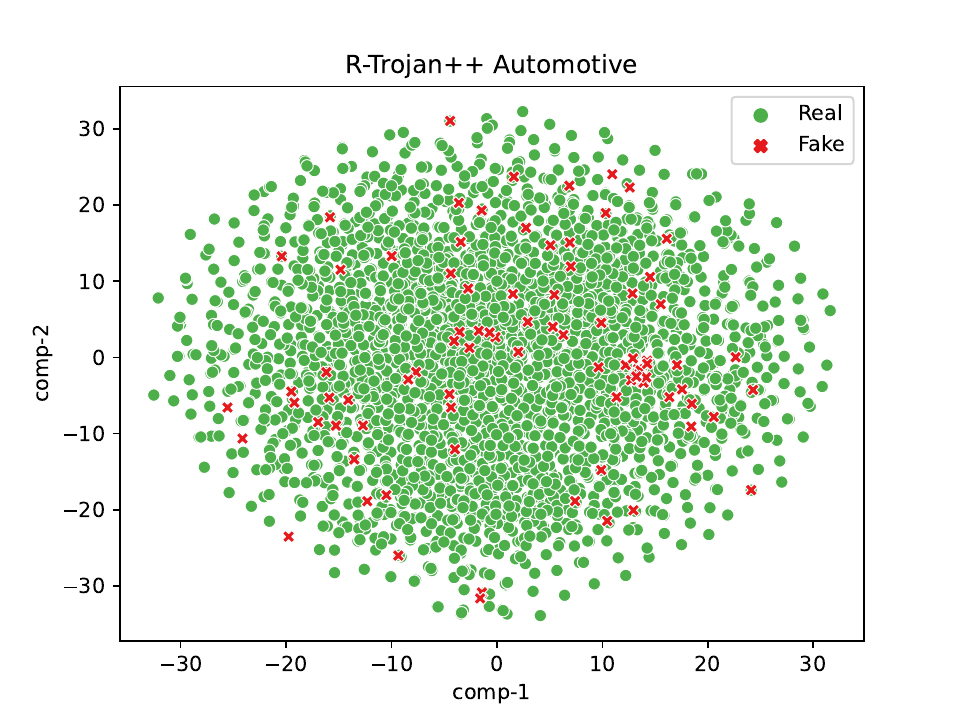}
    \vspace{-1.2em}
    \caption{RAGAN on Automotive}
    \label{SUBFIGURE LABEL 2}
\end{subfigure}
\begin{subfigure}{.24\textwidth}
    \centering
    \includegraphics[width=\linewidth]{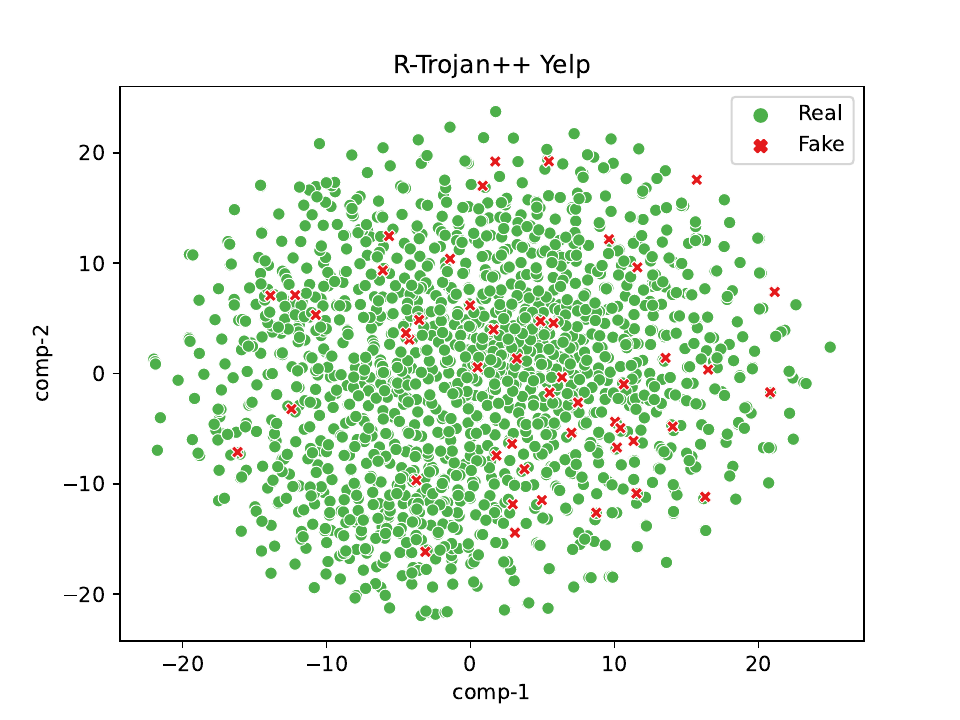}
    \vspace{-1.2em}
    \caption{RAGAN on Yelp}
    \label{SUBFIGURE LABEL 3}
\end{subfigure}
\begin{subfigure}{.24\textwidth}
    \centering
    \includegraphics[width=\linewidth]{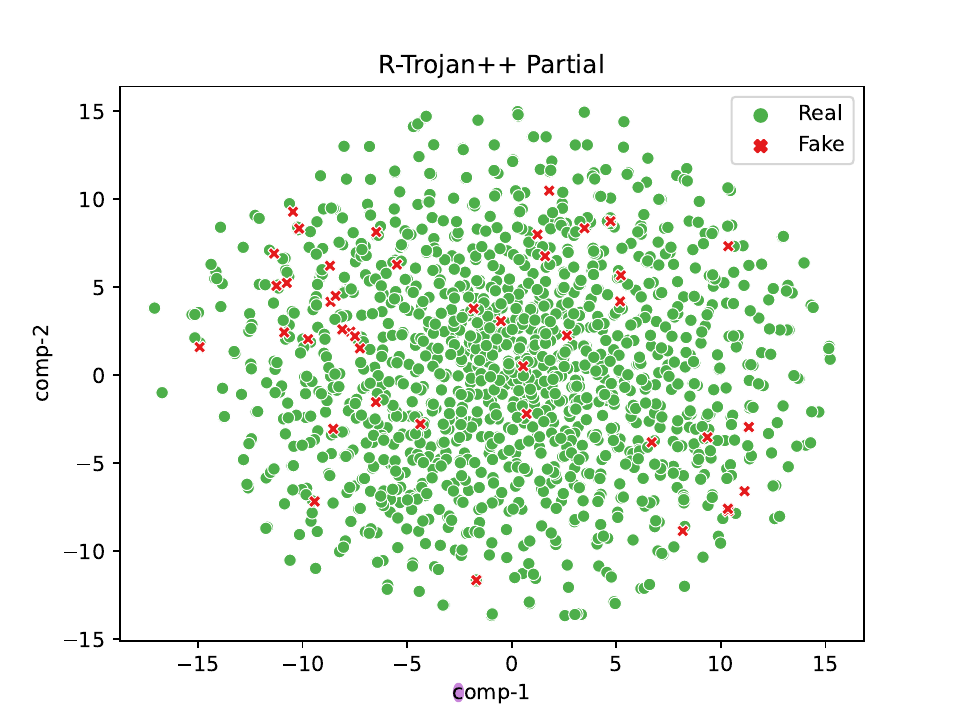}
    \vspace{-1.2em}
    \caption{RAGAN on Partial Musical}
    \label{SUBFIGURE LABEL 4}
\end{subfigure}
\vspace{-0.4em}
\caption{Visualization of RAGAN's fake user profiles and real user profiles on real-world datasets. The attack profiles overlap with normal profiles, reflecting a global semantic alignment and demonstrating the imperceptibility of RAGAN.}
\label{fig:tsne_graphs}
\vspace{-1em}
\end{figure*}

\begin{figure}
\centering
\includegraphics[width=0.9\linewidth]{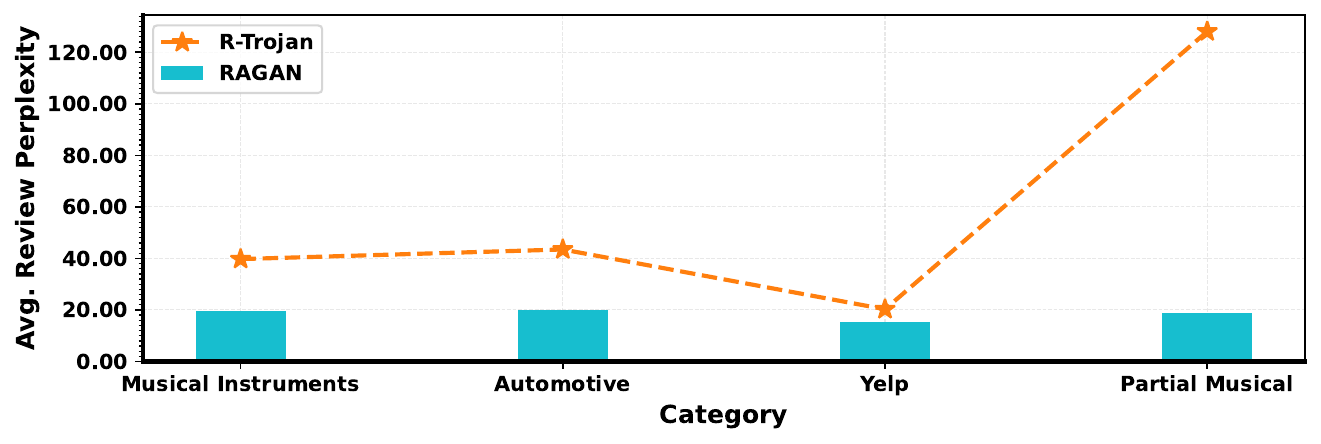}  
\vspace{-0.6em}
\caption{Average text perplexity of  generated reviews from different attacks with varying knowledge on real-world datasets.}
\label{fig:text_perplexity}
\vspace{-1.8em}
\end{figure}

\section{Experiment}

\subsection{Experiment Setup}
\subsubsection{Datasets Selection}
We use three widely-used real-world datasets from different scenarios  \cite{mcauley2015image, yang2025drunkagent} to evaluate RAGAN, which are Amazon Musical Instruments, Amazon Automotive, and Yelp, as detailed in Table \ref{tab:statistics}. For Yelp\footnote{https://www.yelp.com/dataset}, we randomly select a subset to avoid exceeding the hardware limit.
The datasets vary in size and sparsity, which is suitable for a comprehensive evaluation of RAGAN.
We adopt the leave-one-out method \cite{he2017neural} to select the training set and the test set, with a split ratio of 9:1.


\subsubsection{Baseline Attack Methods}
Following the latest work \cite{yang2023incorporated}, we compare RAGAN with a series of representative poisoning attacks including conventional: Random and Bandwagon \cite{gunes2014shilling}, algorithm-specific: PGA \cite{li2016data} and DLA \cite{huang2021data}, and GAN-based: DCGAN \cite{christakopoulou2019adversarial}, AUSH \cite{lin2020attacking}, RecUP \cite{zhang2021attacking}, Leg-UP \cite{lin2022shilling}, TrialAttack \cite{wu2021triple} and state-of-the-art R-Trojan \cite{yang2023incorporated}. To maintain a fair comparison, for all the attack methods, we set the \textit{attack size} to 3\% of the population as default, which can clearly show the performance differences \cite{lin2022shilling}, and the \textit{profile size} equals the average number of ratings per user in the data set. Moreover, to further ensure fair and consistent evaluations, all attacks are optimized using our surrogate agent to avoid high query costs while maintaining stealthiness under practical black-box settings. For reproducibility, the hyper-parameters of the baselines are set as suggested in the original papers.

\subsubsection{Targeted Recommender Systems}
We consider three representative rating-only-based victim RSs, as existing works do \cite{lin2022shilling, lin2020attacking, yang2023incorporated}: WRMF \cite{hu2008collaborative}, NCF \cite{he2017neural} and LightGCN \cite{he2020lightgcn}. Moreover, to comprehensively evaluate the attack effectiveness on review-based RSs, we use typical DeepCoNN \cite{zheng2017joint} and the state-of-the-art agent model \cite{yang2023incorporated} as victim RSs.
For reproducibility, the hyper-parameters of the victim RSs are set as suggested in the original papers.



\subsubsection{Evaluation Metrics}
Following the existing works \cite{zhang2021attacking, lin2020attacking, huang2021data, lin2022shilling, wu2021triple}, we use two widely-used ranking metrics to evaluate attack transferability: hit ratio (HR@$k\uparrow$) and normalized discounted cumulative gain (NDCG@$k\uparrow$), where HR@$k$ measures whether the target item is appeared in the top-$k$ recommendation list, while NDCG@$k$ indicates the ranking position of the target item in the list \cite{wu2021ready}. $k$ is set to 10 here. We further introduce the standard sentence perplexity score $\downarrow$ \cite{zhang2024stealthy}, a commonly-adopted metric for evaluating attack imperceptibility, as textual reviews primarily serve to enhance the stealthiness of the generation profiles \cite{yang2023incorporated}. Perplexity measures the average inverse likelihood of each token under a language model, reflecting how fluent and natural a text appears. Notably, perplexity has also been leveraged in LLM-based detectors for identifying prompt injection attacks \cite{liu2024formalizing}, reinforcing its reliability as a measure of text imperceptibility.

\subsubsection{Configuration of RAGAN}

We set $l$=6, the number of in-context examples to 3, training epochs to 20, batch size to 256, learning rate to 0.001, $\lambda$ to 0.5 to balance the two attack goals. To improve the attack efficiency while reducing the attack costs, we adopt `gemini-1.5-flash' for $f_{\text{FM}_{\zeta}} (\cdot)$, `vit\_large\_patch14\_224\_clip\_laion2b' for CLIP-ViT and `all-MiniLM-L6-v2' for SBERT, where the hyper-parameters are configured as the default values provided by Hugging Face\footnote{https://github.com/huggingface/transformers}. For other hyper-parameters and the data pre-processing scheme, we follow the previous work R-Trojan \cite{yang2023incorporated} to ensure the effectiveness of the evaluation.

\subsection{Attack Transferability}

\subsubsection{Overall Transferability}

Table~\ref{tab:comp_results} and Table~\ref{tab:review_results} illustrate the impressive transferability of RAGAN, underscoring the critical role of high-quality textual reviews in strengthening the effectiveness of the attack profiles across diverse victim RSs. From Table~\ref{tab:comp_results}, we can find that LightGCN is more robust than NCF and WRMF, possibly due to its unique graph convolutions. For the review-based RSs, as shown in Table~\ref{tab:review_results}, the review-based agent is more vulnerable than DeepCoNN, mainly because RAGAN is optimized on an instructional agent with similar architectures and behavioral patterns.
Furthermore, the results on Yelp exhibit less variability, which we attribute to its larger scale and higher density, making it more resistant to the perturbations.

\subsubsection{RAGAN vs. Attack Baselines}

As shown in Table~\ref{tab:comp_results} and Table~\ref{tab:review_results}, RAGAN achieves the state-of-the-art performance  among all evaluated methods against various victim RSs on real-world datasets, with the highest HR@10 and NDCG@10. By refining the quality of textual reviews, RAGAN demonstrates improved attack transferability compared to R-Trojan. R-Trojan greatly outperforms other baselines, with its key distinction being the incorporation of textual reviews, demonstrating that even basic utilization of the reviews can significantly improve the attack performance. 

Among the remaining baselines, Leg-UP, TrialAttack and DLA exhibit comparable performance, likely due to their use of surrogate RSs (Leg-UP and TrialAttack), which enhance attack transferability, and DLA’s optimization for DL-based RSs. RecUP and AUSH achieve moderate performance, outperforming DCGAN due to their specialized GAN architectures for shilling attacks. In contrast, PGA, which is tailored for MF models, fails to effectively transfer to DL-based RSs such as NCF and LightGCN. Finally, conventional attack methods (i.e., Random and Bandwagon) demonstrate the lowest transferability among all baselines.

\begin{figure}
\centering
\begin{subfigure}{.3\textwidth}
    \centering
    \includegraphics[width=\linewidth]{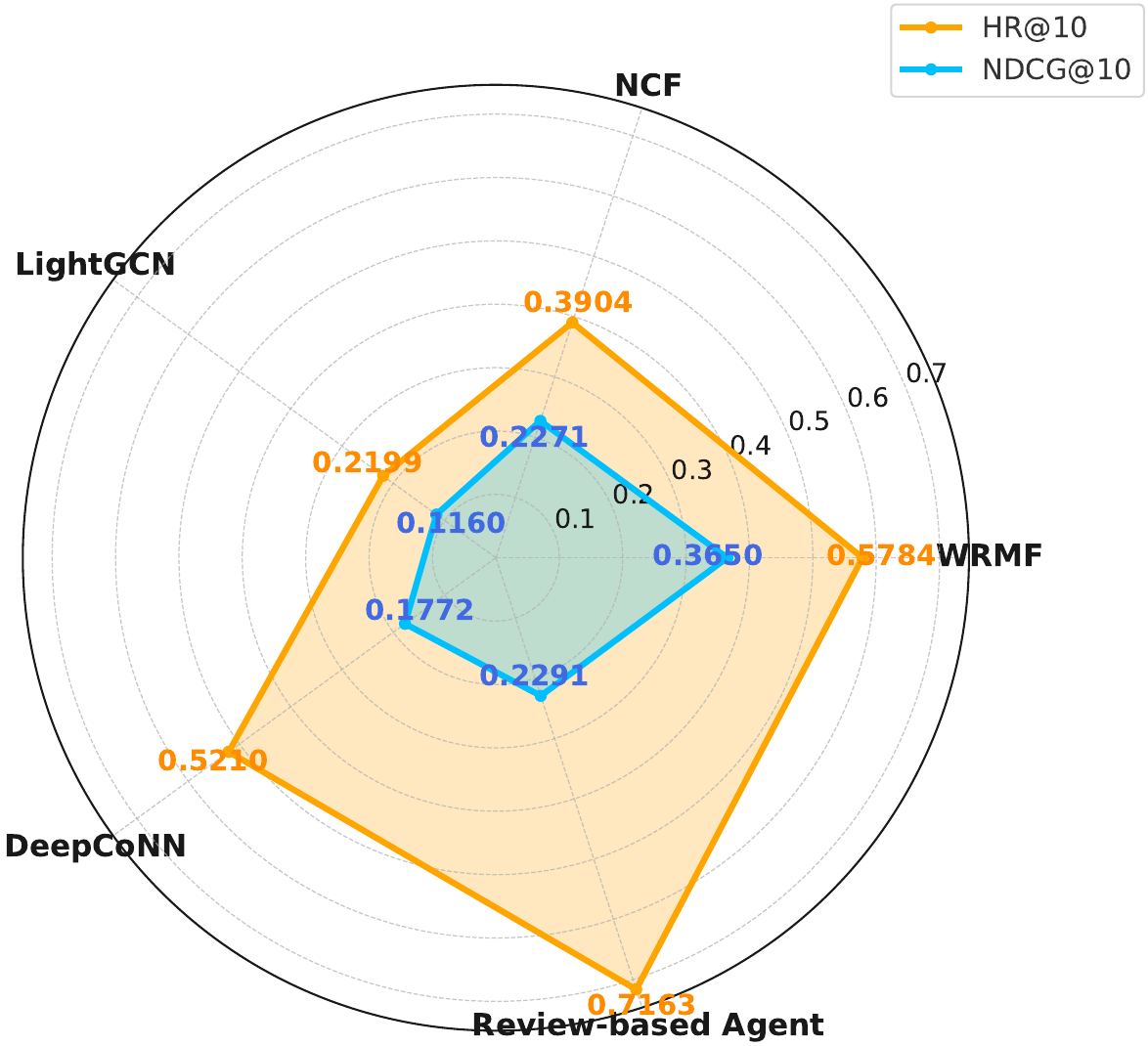}
    \vspace{-1.2em}
    \caption{RAGAN on Amazon}
    \label{radar_amazon}
\end{subfigure}
\begin{subfigure}{.3\textwidth}
    \centering
     \includegraphics[width=\linewidth]{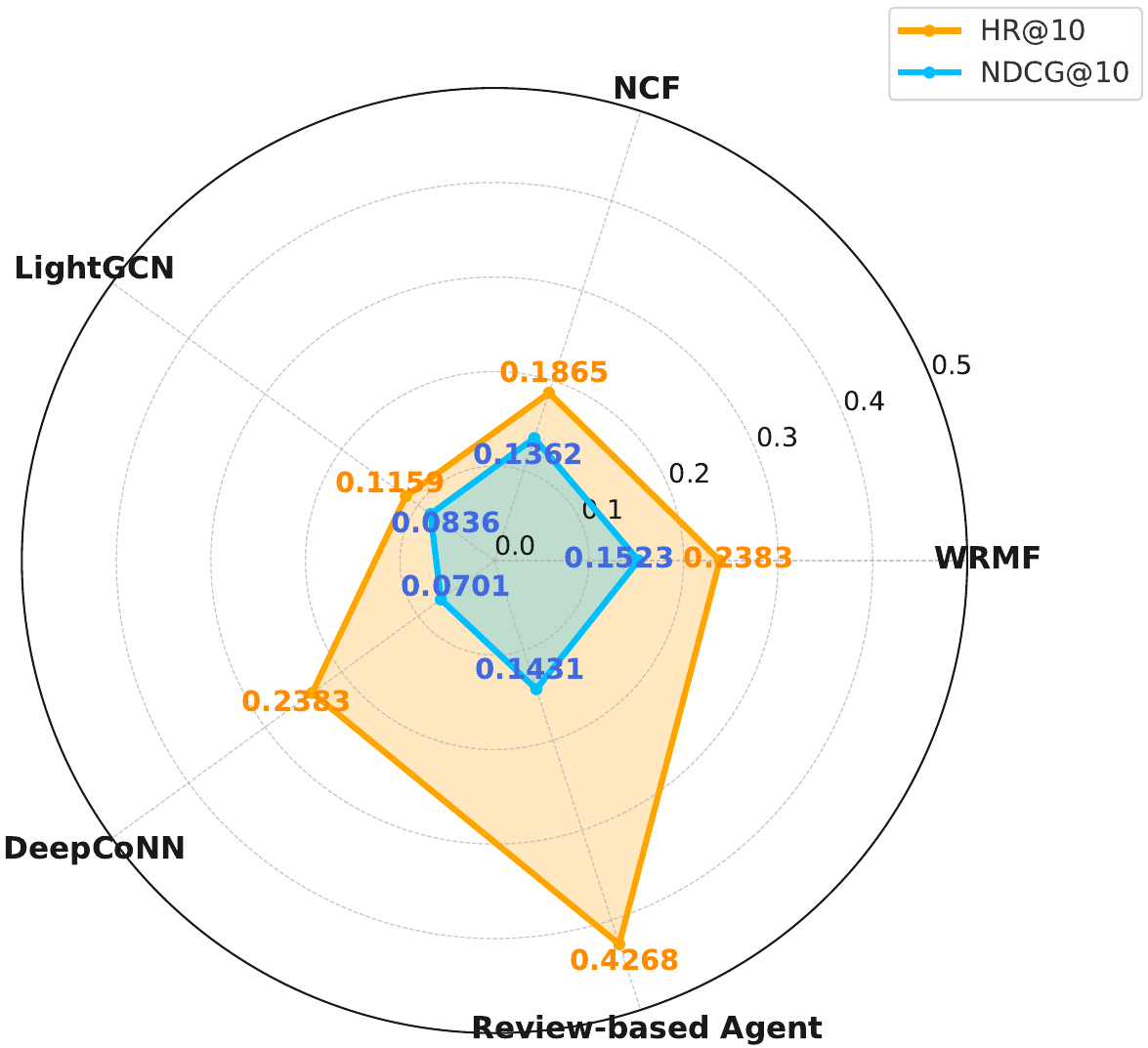}
    \vspace{-1.2em}
    \caption{RAGAN on Yelp}
    \label{radar_yelp}
\end{subfigure}
\vspace{-0.4em}
\caption{The transferability of RAGAN with partial knowledge against various black-box RSs on real-world datasets.}
\label{fig:radar_charts}
\vspace{-1.5em}
\end{figure}

\subsection{Attack Imperceptibility}

\subsubsection{Overall Imperceptibility}

Since we are very early attempts to incorporate textual reviews into fake user profiles to enhance attack effectiveness, existing detectors, which currently focus on rating-only profiles (e.g., \cite{zeng2023practical, lin2022shilling}), have semantic gaps for identifying RAGAN. To demonstrate the imperceptibility of the attack, we employ t-SNE \cite{lin2022shilling}, which is a typical method used in the current works \cite{yang2023incorporated, yang2024attacking}, to visualize the real user's and RAGAN's profile representations generated by the guardian module on real-world datasets, which involves rating and review information. As shown in Fig.~\ref{fig:tsne_graphs}, the representations of fake user profiles are scattered within the distribution of real user profiles on real-world datasets. The overall proximity between attack profiles and normal profiles in the embedding space can be attributed to the imperceptibility optimizations, which enforces high semantic similarity. Meanwhile, the dispersed nature of the attack profiles may stem from our style transfer strategy, which enhances the diversity and realism of the generated behaviors. Hence, RAGAN can effectively mimic real users, enabling highly inconspicuous attacks.

\begin{table*}
  \caption{Ablation studies of RAGAN on real-world datasets.}
  \vspace{-1em}
  \begin{center}
  \small
  \label{tab:ablation_study}
  \resizebox{2\columnwidth}{!}{
  \begin{tabular}{c|c|cc|cc|cc|cc|cc}
    \hline

    \multirow{2}{*}{Dataset}  & \multicolumn{1}{c|}{\multirow{2}{*}{Component}} & \multicolumn{2}{c|}{WRMF} & \multicolumn{2}{c|}{NCF} & \multicolumn{2}{c|}{LightGCN} & \multicolumn{2}{c|}{DeepCoNN} & \multicolumn{2}{c}{Review-based Agent} \\

    \cline{3-12} 
    
    & & HR@10 & \multicolumn{1}{c|}{NDCG@10} & HR@10 & \multicolumn{1}{c|}{NDCG@10} & HR@10 & \multicolumn{1}{c|}{NDCG@10}  &  HR@10 & \multicolumn{1}{c|}{NDCG@10} & HR@10 & NDCG@10\\

    \hline
     \multirow{3}{*}{Amazon} & RAGAN$^a$ & 0.5581 & \multicolumn{1}{c|}{0.3137} & 0.3527 & \multicolumn{1}{c|}{0.2078} &  0.1967 & \multicolumn{1}{c|}{0.1079} & 0.4064 &\multicolumn{1}{c|}{0.1191} & 0.6081 & 0.1901 \\

     & RAGAN$^b$  & 0.4673 &\multicolumn{1}{c|}{0.2619} &  0.2605 & \multicolumn{1}{c|}{0.1316} & 0.1407 & \multicolumn{1}{c|}{0.0741} & 0.3019 & \multicolumn{1}{c|}{0.0882} & 0.4369 & 0.1316 \\


      & RAGAN &  \textbf{0.6313} & \multicolumn{1}{c|}{\textbf{0.4254}} &  \textbf{0.4209} & \multicolumn{1}{c|}{\textbf{0.2526}} &  \textbf{0.2569} & \multicolumn{1}{c|}{\textbf{0.1331}} & \textbf{0.5624}  &\multicolumn{1}{c|}{\textbf{0.2018}} & \textbf{0.7438}  & \textbf{0.2334} \\

    \hline

    \multirow{3}{*}{Yelp} & RAGAN$^a$ & 0.1859 & \multicolumn{1}{c|}{0.1340} &  0.1593 & \multicolumn{1}{c|}{0.0750} & 0.1023  &\multicolumn{1}{c|}{0.0650} & 0.2267 &\multicolumn{1}{c|}{0.0666} & 0.3303 & 0.1099\\

      & RAGAN$^b$ &  0.1276 &\multicolumn{1}{c|}{0.0938} & 0.1250 & \multicolumn{1}{c|}{0.0659} &  0.0654 & \multicolumn{1}{c|}{0.0459} & 0.0848 & \multicolumn{1}{c|}{0.0248} & 0.1587 & 0.0468 \\


      & RAGAN &  \textbf{0.2494} & \multicolumn{1}{c|}{\textbf{0.2011}} & \textbf{0.2040} & \multicolumn{1}{c|}{\textbf{0.1469}} & \textbf{0.1308}  & \multicolumn{1}{c|}{\textbf{0.0961}} & \textbf{0.2597} & \multicolumn{1}{c|}{\textbf{0.0760}} & \textbf{0.4495} & \textbf{0.1394} \\

   \hline
    
  \end{tabular}}
  \end{center}
\vspace{-1.5em}
\end{table*}

\begin{table*}
  \caption{Examples of fake ratings with the corresponding fake textual reviews from different attacks on real-world datasets. }
  \vspace{-1em}
  \begin{center}
  \small
\label{tab:profile_exp}
\resizebox{2\columnwidth}{!}{\begin{tabular}{c|c|c|c}
\hline
  Dataset & Attack & Fake Review Text Generated by Attack Methods & Fake Rating \\
  \hline
  
  \multirow{7}{*}{Amazon} & \multirow{2}{*}{R-Trojan} & \textbf{Excellent Guitar Stand Black} I've had several stands for my electric guitars and none of them had a black coating on them. The only thing I wanted & \multirow{2}{*}{5.0} \\
  & &  was black. A non-skid coat like Fender, suede or something would have been perfect. I'm glad I did. & \\

  \cline{2-4}

   & \multirow{5}{*}{RAGAN}  & \textbf{This Musician's Gear Tubular Guitar Stand} is a lifesaver!  At my age, getting my guitar in and out of its case can be a bit of a struggle.  This stand  & \multirow{5}{*}{5.0} \\ 
   & &  is a perfect solution. It's so easy to use; just set the guitar in place—the soft rubber tubing is gentle on the instrument, and the neck rest  keeps it sec-  & \\ 
   & & ure. The stand is very stable, and the sturdy, wide base gives me peace of mind that my guitar won't tip over. I also appreciate how easily it folds up & \\ 
   &  &   for storage –  it’s wonderfully compact when not in use.  It's simple to set up and put away, no complicated assembly required. For the price, it's a fa-  & \\
   & &  ntastic value. I highly recommend it to anyone, especially those of us who value ease of  use and a secure place to keep their prized instrument. & \\
   \hline
   
  \multirow{11}{*}{Yelp}  & \multirow{5}{*}{R-Trojan} & \textbf{Average RA Sushi Bar Restaurant} My SO and I visited RA Sushi for happy hour (Saturday) around 6:30.  The happy hour prices were very reasonable, & \multirow{5}{*}{3.0}\\
  
  & & including 1/2 price beer, and the sushi was pretty great.  It was the best sushi we've found in the area.  We ordered the tuna poke, the Hawaiian, and the & \\
  
  & & Godzilla.  The Hawaiian was amazing - I could have easily made this my meal if I didn't have to.  The Hawaii was good, but was also a lot better - mu-   & \\
  
  & & ch more flavorful and fresh. The sushi was pretty fresh and it looked like there was really really nothing wrong with it.  It was only \$5 cheaper, but I d- &  \\

  & &  on't  know if we would have gone there for dinner. I would definitely go back. & \\

  \cline{2-4}
  
  & \multirow{6}{*}{RAGAN}  & My recent visit to \textbf{RA Sushi} was a bit of a mixed bag.   The restaurant itself is pleasant enough, easy to navigate, and the seating was comfortable.  I   & \multirow{6}{*}{3.0} \\ 

  &  & appreciated the large print  on the menu, making it simple to read.    My order, a simple California roll and miso soup, was accurately taken.   The soup & \\

  &  & was   delicious, perfectly warm and flavorful. However, the sushi rice was  a little bit on the dry side for my taste. The service was friendly enough, bu- & \\

   &  & t a little slow;   we had to ask  for our check twice. Overall, it was a decent experience,  but nothing exceptional. The food was good, but not outstandin-  & \\

   &  & g, and the service   was adequate  but not overly attentive.  For a casual meal,  it’s acceptable,  but I wouldn't rush back, especially if you're looking for & \\

   &  & particularly speedy service.     The bill was reasonable. & \\
  
  \hline
  \end{tabular}}
  \end{center}
  \vspace{-2.3em}
\end{table*}

\subsubsection{Review Quality}

To further assess the imperceptibility of the attack, we conduct an additional evaluation on the generation quality of the adversarial reviews. 
Following the works \cite{zhang2024stealthy, yang2025drunkagent}, we adopt the text perplexity score computed by GPT-Neo as the metric. A lower perplexity indicates that the perturbed text remains closer to natural human language, thus appearing less suspicious and undetectable. As shown in Fig.~\ref{fig:text_perplexity}, RAGAN achieves a lower average perplexity than R-Trojan, demonstrating its superiority in generating high-quality review text. These reviews are not only fluent and coherent but also semantically meaningful and stealthy, thereby enhancing their deceptive effectiveness. This performance advantage arises because RAGAN effectively harnesses the full potential of its multimodal FMs powered by the retrieval demonstration algorithm and the text style transfer strategy, thereby achieving stronger ICL and generative capabilities than R-Trojan’s plain prompting approach.

\subsection{Attack Performance with Partial Knowledge}
To  evaluate the effectiveness of RAGAN in more practical settings, we conduct experiments with the assumption that the attacker can access only 50\% of the user-item interaction records, i.e., $p=0.5$. Due to space limitations, we focus on representative datasets, reporting mainly on Amazon Musical Instruments and additionally on Yelp for a more comprehensive evaluation. For the other datasets, the trends are similar. From Fig.~\ref{fig:radar_charts} and Fig.~\ref{fig:tsne_graphs}(d), we can observe that RAGAN is still effective against various black-box RSs on real-world datasets, and its profiles remain similar to normal user profiles, with high-quality reviews generated. These results indicate that RAGAN can generate transferable and imperceptible fake user profiles, even when only partial knowledge of the training data of the victim RSs is accessed. This weak model in Fig.~\ref{fig:radar_charts} is significantly superior to all the baselines learned with the full data in Table~\ref{tab:comp_results} and Table~\ref{tab:review_results}, and the same advantage is also shown in Fig.~\ref{fig:text_perplexity}, demonstrating the effectiveness of the semantically rich textual reviews in reinforcing the attack. These findings highlight that existing RSs are vulnerable to semantically informed threat models in realistic scenarios.

\subsection{Ablation Studies}

We remove some pivotal components of RAGAN and investigate the attack performance changes. Table~\ref{tab:ablation_study} illustrates the performance of ablation studies of RAGAN. We give results on two representative datasets, which are Amazon Musical Instruments and Yelp, due to space limitations.

To validate the effectiveness of our review quality enhancement component, we conduct ablation experiments by removing the multimodal demonstration retrieval algorithm (denoted as RAGAN$^a$). Since the previous work \cite{yang2023incorporated} has empirically demonstrated the effectiveness of textual reviews for strengthening fake user profiles, we focus solely on assessing how review quality affects attack performance. By comparing the cases of RAGAN$^a$ (zero-shot) and the full RAGAN (few-shot), we observe that improving the quality of fake textual reviews contributes notably to the transferability of the attack. The main reason may be that the reviews encapsulate rich item features, which can help focus on a broader set of users with similar taste preferences and thus enhance item exposure rates. Moreover, as shown in Table~\ref{tab:comp_results}, Table~\ref{tab:review_results} and Table~\ref{tab:ablation_study}, RAGAN$^a$ achieves comparable performance to R-Trojan across various victim RSs. This is primarily attributed to the task-specific prompting techniques tailored on the powerful FMs, which enhances the models' capabilities in understanding, reasoning and generation, thereby producing more semantically meaningful review text than the simple prompting method used in R-Trojan. Moreover, removing the instructional agent (denoted as RAGAN$^b$) results in a significant drop in attack performance against different victim models, as shown in Table~\ref{tab:ablation_study}, indicating its vital role in optimizing and improving the transferability of the attack.

We do not ablate the Guardian module, as removing it would break the GAN framework and make the adversarial optimization meaningless. Similarly, the style transfer strategy is kept unchanged, because it is a prompt-level cue designed to add linguistic diversity for more natural review expressions rather than a learnable module. Empirically, its removal yields negligible changes in attack effectiveness.


\subsection{Examples of Adversarial Attack Profiles}

We provide examples to demonstrate the quality of the attack profiles generated by R-Trojan and RAGAN (note that R-Trojan is the first and only method to incorporate reviews into the profiles in the literature). Due to space constraints, we randomly select one item from the profiles on each type of dataset. As illustrated in Table~\ref{tab:profile_exp}, the reviews in R-Trojan are incoherent, factually inaccurate,   semantically confused, logically inconsistent, sparse in content and misaligned with the ratings. In contrast, the reviews in RAGAN are more semantically meaningful, richer in valuable features, more coherent and natural, and better aligned with the ratings, resulting in higher-quality profiles. This observation also sheds light on why RAGAN is effective and why its profiles are both transferable and imperceptible.

\section{Conclusion and Future Works}

Many existing attack methods are overly idealized and fail to reflect realistic adversarial scenarios, which can result in overly optimistic assessments of the robustness of RSs in practice. To mitigate this problem, in this paper, we propose a novel poisoning attack framework, RAGAN, for generating high-quality fake user profiles including carefully designed ratings and textual reviews. RAGAN achieves significant improvements over existing profiles, particularly in the coherence and semantic richness of the reviews, as well as in the alignment between ratings and reviews. As a result, RAGAN demonstrates state-of-the-art transferability and imperceptibility in a more practical setting, as shown in the comprehensive experiments on real-world datasets. RAGAN enables us to delve into the vulnerabilities of current RSs, revealing their inherent security risks under specific input features (e.g., well-crafted ratings and/or reviews), and guides us to secure them in actual deployment scenarios. Since our works are conducted on real-world datasets, future work will extend the RAGAN framework to real-world RSs for realistic robustness analysis and systematic defense testing, paving the way toward more secure practical deployments. Beyond this, we plan to 1) explore adaptive defense mechanisms that dynamically counteract evolving poisoning strategies, and 2) generalize RAGAN to multimodal and LLM-based recommenders, where textual, visual, and conversational signals jointly influence recommendations.

\appendix
\section*{Text Style Transfer Prompt Corpus} \label{text_style}
To support nuanced and controllable text style transfer, we construct a comprehensive prompt corpus organized into ten major categories: \textit{sentiment, formality, persona, emotion, complexity, audience, creativity, comparison, genre, and dialect}. Each category captures a distinct dimension of stylistic variation. For instance, sentiment-based prompts guide the tone of opinion expression (e.g., positive, negative, or neutral), while formality-based prompts regulate the register from casual to highly formal. Persona-based prompts simulate different user perspectives such as experts, first-time buyers, parents, or teenagers. Emotion-based prompts capture a wide range of affective tones including joy, disappointment, anger, fear, and gratitude. Complexity-based prompts control linguistic sophistication, ranging from plain, accessible language to technical or complex styles. Audience-specific prompts adapt the review for particular demographic groups like families, professionals, students, or seniors. Creativity-focused prompts encourage stylistic diversity through humor, satire, storytelling, or poetic forms. Comparative prompts introduce contrastive framing, such as  before-and-after contrasts or upgrades over time. Genre-based prompts reframe content into different communicative styles, including advertisement, news report, dialogue, or social media post. Finally, dialect prompts adjust for linguistic variation across English dialects (e.g., British vs. American) and regional colloquialisms to reflect diverse cultural norms. Hence, this taxonomy enables fine-grained control over stylistic modifications.

\bibliographystyle{IEEEtran}
\bibliography{IEEEabrv, IEEEexample}

\end{document}